\documentclass[12pt,sort&compress]{article}
\usepackage{graphicx}% Include figure files
\usepackage{bm}% bold math
\usepackage{amsmath,amssymb,color,mathrsfs}
\usepackage{footnote}
\usepackage{citesort}% compacts and sorts references
\usepackage{cuted}

\newcommand{\PR}{Phys. Rev. }
\newcommand{\PRL}{Phys. Rev. Lett. }
\newcommand{\PRA}{Phys. Rev. A }
\newcommand{\PRB}{Phys. Rev. B }

\newcommand{\PRE}{Phys. Rev. E }
\newcommand{\PLA}{Phys. Lett. A }
\newcommand{\APL}{Appl. Phys. Lett. }

\newcommand{\jpa}{J. Phys. A: Math. Theor. }
\newcommand{\JPA}{J. Phys. A: Math. Gen. }
\newcommand{\AJP}{Am. J. Phys. }
\newcommand{\AJM}{Am. J. Math. }

\newcommand{\PA}{Physica A }
\newcommand{\CMP}{Commun. Math. Phys. }
\newcommand{\AM}{Annals Math. }
\newcommand{\APB}{Ann. Phys. (Berlin) }
\newcommand{\APNY}{Ann. Phys. (N. Y.) }

\definecolor{officegreen}{rgb}{0,0.5,0}
\definecolor{pakistangreen}{rgb}{0,0.4,0}
\definecolor{palatinatepurple}{rgb}{0.41,0.16,0.38}
\definecolor{sangria}{rgb}{0.57,0,0.04}

\begin{document}
\title{Comparative analysis of electric field influence on the quantum wells with different boundary conditions. I. Energy spectrum, quantum information entropy and polarization}
\author{O. Olendski\footnote{King Abdullah Institute for Nanotechnology, King Saud University, P.O. Box 2455, Riyadh 11451 Saudi Arabia; E-mail: oolendski@ksu.edu.sa}}

\maketitle

\begin{abstract}
Analytical solutions of the Schr\"{o}dinger equation for the one-dimensional quantum well with all possible permutations of the Dirichlet and Neumann boundary conditions (BCs) in perpendicular to the interfaces uniform electric field $\mathscr{E}$ are used for the comparative investigation of their interaction and its influence on the properties of the system. Limiting cases of the weak and strong voltages allow an easy mathematical treatment and its clear physical explanation; in particular, for the small $\mathscr{E}$, the perturbation theory derives for all geometries a linear dependence of the polarization on the field with the BC-dependent proportionality coefficient being positive (negative) for the ground (excited) states. Simple two-level approximation elementary explains the negative polarizations as a result of the field-induced destructive interference of the unperturbed modes and shows that in this case the admixture of only the neighboring states plays a dominant role. Different magnitudes of the polarization for different BCs in this regime are explained physically and confirmed numerically. Hellmann-Feynman theorem reveals a fundamental relation between the polarization and the speed of the energy change with the field. It is proved that zero-voltage position entropies $S_x$ are BC independent and for all states but the ground Neumann level (which has $S_x=0$) are equal to $\ln2-1$ while the momentum entropies $S_k$ depend on the edge requirements and the level. Varying electric field changes position and momentum entropies in the opposite directions such that the entropic uncertainty relation is satisfied. Other physical quantities such as the BC-dependent zero-energy and zero-polarization fields are also studied both numerically and analytically. Applications to different branches of physics, such as ocean fluid dynamics and atmospheric and metallic waveguide electrodynamics, are discussed.
\end{abstract}

\section{Introduction}\label{sec_Intro}
One-dimensional (1D) quantum well (QW) with vanishing at its edges wavefunction $\Psi(x)$ is an example that is included in every textbook on quantum mechanics \cite{Landau1}.  Influence on its properties of an electric field that is applied perpendicularly to the confining planes \cite{Rabinovitch1,Lukes1,Fernandez1,Bastard1,Ahn1,Nguyen1} is of not only purely academic interest but also of the large technological applications. By creating the chosen direction, the field lowers the overall symmetry of the structure what results in remarkable phenomena such as, for example, the change in the interband optical electroabsorption of a semiconductor QW as compared to the bulk samples. This quantum-confined Franz-Keldysh effect \cite{Miller1,Miller2,Weiner1,Miller3,Achtstein1} is employed in the design of the devices of integrated optics, such as high-speed modulators \cite{Wood1},  optically bistable switches \cite{Miller4} and wavelength selective detectors \cite{Wood2}.

The condition
\begin{equation}\label{Dirichlet1}
\Psi|_{\cal S}=0
\end{equation}
of the zeroing at the confining surface $\cal S$ of the solution $\Psi$ of the wave equation is not the only demand implemented in the analysis of the physical and chemical systems. Another frequently used type of the boundary condition (BC) is the requirement of zeroing at the interface of the normal derivative of $\Psi$:
\begin{equation}\label{Neumann1}
\left.\frac{\partial\Psi}{\partial {\bf n}}\right|_{\cal S}=0.
\end{equation}
Here, $\bf n$ is an inward unit normal to the surface. This condition is satisfied, for example, by the transverse electric waves in the electromagnetic waveguides \cite{Jackson1} or the order parameter of the superconductor bordering the vacuum or the insulator \cite{DeGennes1}. The planar structures with different BCs on the opposite confining walls \cite{Olendski1} are also encountered in nature; for example, the long-distance dynamics between the Earth and ionosphere of the very low frequency electromagnetic oscillations with wavelengths $10\lesssim\lambda\lesssim100$ km is described, in the first approximation, as a propagation in the flat waveguide with one of its walls (the ground) being a perfect electric conductor with a reflection coefficient $+1$ what corresponds to the BC from Eq.~\eqref{Neumann1}, while the ionosphere behaves as an ideal magnetic conductor with a reflection coefficient $-1$ with the edge requirement governed by Eq.~\eqref{Dirichlet1} \cite{Davies1}. Similarly, the sound pressure in the acoustical waveguide formed by the air-water surface and the bottom of the lake is zero at the upper free boundary with its vanishing derivative at the pond's rigid floor \cite{Budden1}. The same configuration of the BCs is used, e. g., in the description of the continental shelf waves along curved coasts \cite{Johnson1}. A short review of other structures with mixed combination of the Dirichlet and Neumann BCs, which are called Zaremba geometries \cite{Zaremba1}, is given in Ref.~\cite{Najar1}.

In the present research, exact calculations of the QW with all possible permutations of the BCs from Eqs.~\eqref{Dirichlet1} and \eqref{Neumann1} in the uniform transverse electric field are presented and analyzed with the emphasis on the interrelations between them. While the case of the Dirichlet, Eq.~\eqref{Dirichlet1}, BCs at the both edges is quite well known \cite{Rabinovitch1,Lukes1,Fernandez1,Bastard1,Nguyen1,Ahn1,Miller1,Miller2,Weiner1,Miller3}, other distributions of the interface requirements with a voltage drop between them have received almost no attention \cite{Exner1}. Meanwhile, they are indispensable, for example, in the analysis of the processes in the metallic waveguide with insulator or ferrite inner filling whose refractive index linearly changes between the confining walls. In addition, if the electron concentration in the ionosphere is also a linear function of the height above the Earth \cite{Budden2}, then the corresponding phenomena will be described too by the rules presented below. However, our best intention here is not only to advice specific applications of the obtained results but, first of all, to build up a more general picture, which is also of the large didactic significance, by establishing interdependencies, drawing parallels and  discovering similarities and differences between miscellaneous geometries of the BCs and their interplay with the field. Following purely Dirichlet case \cite{Rabinovitch1,Lukes1,Fernandez1,Bastard1,Nguyen1,Ahn1,Miller1,Miller2,Weiner1,Miller3}, we, for definiteness, will speak about the quantum particle of the mass $m$ and charge $-e$ ( with $e$ being an absolute value of the electron charge) and the wavefunction $\Psi$ that governs its motion through the solution of the corresponding steady state Schr\"{o}dinger equation. A transformation to, e.g., Maxwell equations is elementary to do as both these relations belong to the same class that in mathematics is called a Helmholtz equation. Excellent discussion of the analogies between the Helmholtz and time-independent Schr\"{o}dinger equations \cite{Gaponenko1} shows that drawing such parallels becomes one of the main driving ideas in nanophotonics. Upon calculating and analyzing energies and functions, a derivation is done of the polarizations of the states, which are a quantitative measure of a system response to the field. A fundamental relation that directly follows from the Hellmann-Feynman theorem and links the polarization with the rate of the energy change with the field, is derived and analyzed.  Consideration of the asymptotic limits of the weak and strong fields allows to arrive at simple analytical expressions. Quantum information entropies in position and momentum space are calculated and analyzed too; in particular, an andvantage of the entropic uncertainty relation with respect to its Heisenberg counterpart is shown for the non-Dirichlet BCs. For the zero field, it is proved that the position entropy is BC independent and for all states but the lowest Neumann level equals to $\ln2-1$ while the momentum entropy depends on the level and type of surface requirement. The field varies position and momentum entropies in the opposite directions in such a way that the corresponding uncertainty relation is always satisfied. Some other results, such as the magnitude of the field at which the excited-state polarizations vanish, are discussed too both numerically and analytically. Summary of the properties of the eigenvalues and eigenfunctions of the Laplace operator in flat arbitrary bounded Euclidean domains with, in particular, Dirichlet, Eq.~\eqref{Dirichlet1}, or Neumann, Eq.~\eqref{Neumann1}, BC is given in a recent review, Ref.~\cite{Grebenkov1}. So, in a sense, the discussion below is a first extension of the previous research for the simplest structure with different edge conditions biased by the electric field. A companion paper \cite{Olendski2} uses the results obtained here for the calculation of the statistical properties of the same system.

The outline below is as follows. In Sec.~\ref{sec_Model} our physical model is introduced and the necessary mathematical formalism for its analysis is developed. Sec.~\ref{sec_Results} is devoted to the presentation of the results with their detailed numerical and analytical investigation. Sec.~\ref{sec_Conclusions} wraps up the study by some conclusions.

\section{Model and Formulation}\label{sec_Model}
\begin{figure}
\centering
\includegraphics[width=0.85\columnwidth]{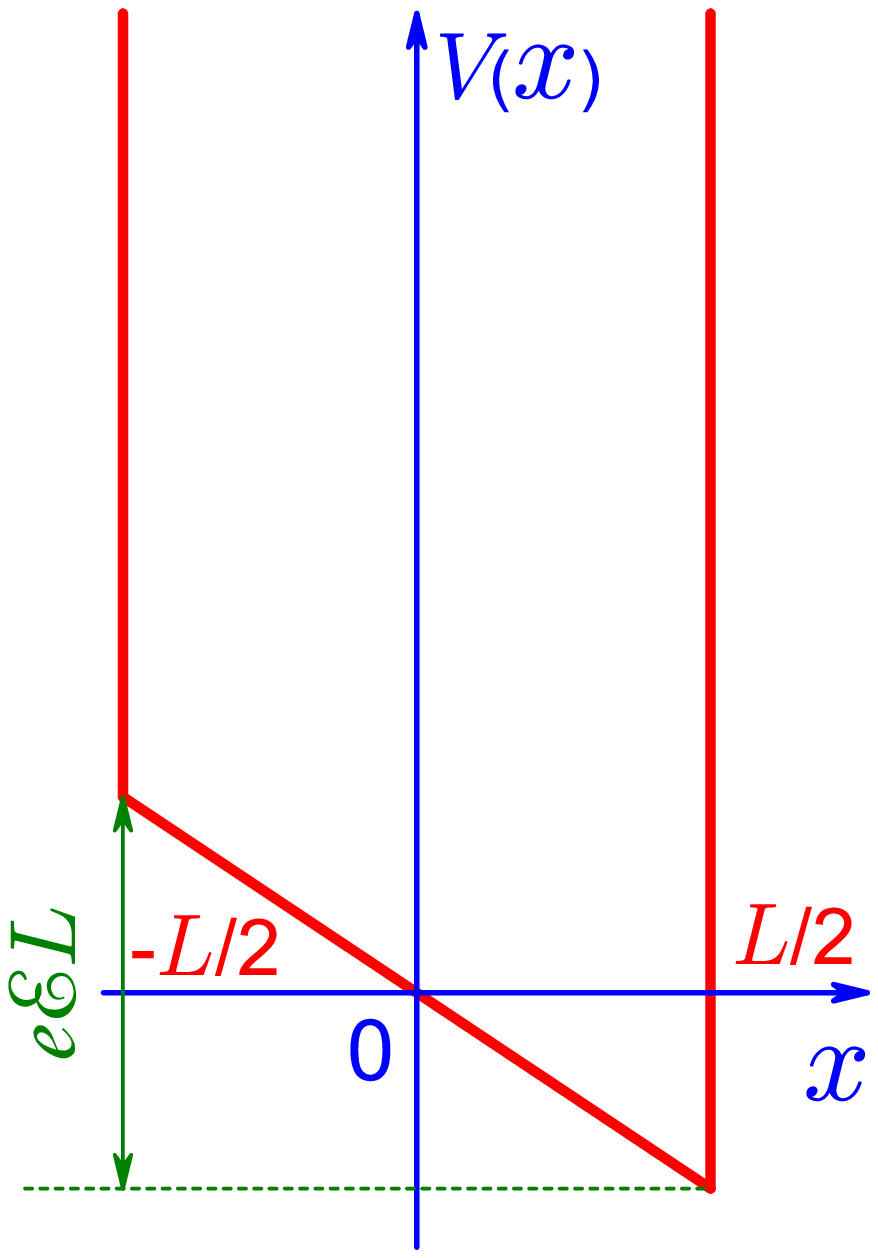}
\caption{\label{Profile}
Potential profile $\color{blue}V(x)$ of the QW located between ${\color{red}-L/2}\le {\color{blue}x}\le{\color{red}L/2}$ in the uniform electric field $\color{pakistangreen}\mathscr{E}$ directed in the negative $\color{blue}x$ direction. Origin of the potential axis coincides with the middle of the potential drop across the well. At each wall either Dirichlet or Neumann BC is satisfied.}
\end{figure}

A charged particle in the 1D infinitely deep QW of the width $L$ is subject to a uniform electric field $\mathscr{E}$ that is applied perpendicularly to its interfaces, Fig.~\ref{Profile}. On each of the walls either Dirichlet (D) or Neumann (N) BC is imposed. Accordingly, below we will  denote the structure by the two characters where the first (second) one refers to the BC at the left (right) wall. Of course, ND configuration can be considered as the DN geometry with the electric field pointing in the opposite direction.  However, to keep the results consistent with the uniform BCs that are symmetric with respect to the sign change of the field, we discuss positive values of $\mathscr{E}$ only and consider ND and DN cases separately. The Hamiltonian $\hat{H}$ in the Schr\"{o}dinger equation
\begin{equation}\label{Schrodinger1}
\hat{H}\Psi(x)=E\Psi(x)
\end{equation}
consists of the field-free part $\hat{H}_0$ and the electrostatic energy $V_\mathscr{E}(x)$ of the field
\begin{equation}\label{Hamilton1}
\hat{H}=\hat{H}_0+V_\mathscr{E}(x)
\end{equation}
with
\begin{equation}\label{Hamilton0}
\hat{H}_0=-\frac{\hbar^2}{2m}\frac{d^2}{dx^2}+V_{QW}(x)
\end{equation}
and
\begin{equation}\label{V1}
V_\mathscr{E}(x)=-e\mathscr{E}x.
\end{equation}
QW potential $V_{QW}(x)$ reads:
\begin{equation}\label{PotentialQW1}
V_{QW}(x)=\left\{
\begin{array}{cc}
0,&|x|<L/2\\
\infty,&|x|\geq L/2.
\end{array}
\right.
\end{equation}
Our primary goal is to find and analyze energies $E$ and wavefunctions $\Psi(x)$ for all possible permutations of the BCs and, based on them, to calculate other quantities describing the system; in particular, polarization $P$ that is defined as \cite{Nguyen1}
\begin{equation}\label{Polarization1}
P(\mathscr{E})=\left<ex\right>_\mathscr{E}-\left<ex\right>_{\mathscr{E}=0}
\end{equation}
with $\left<\ldots\right>$ denoting a quantum mechanical expectation value:
\begin{equation}\label{QMexpectation1}
\left<x\right>=\int_{-L/2}^{L/2}x\Psi^2(x)dx.
\end{equation}

It is convenient from the very beginning to use dimensionless units such that the distances are measured in units of the well width $L$, and energies  - in units of the ground state energy $\pi^2\hbar^2/(2mL^2)$ of the Dirichlet QW. Then, the unit of the electric field $\pi^2\hbar^2/(2emL^3)$ is naturally determined from the voltage drop $e\mathscr{E}L$ across the well, and unit of the polarization will be $eL$. Hence, the field-free part $\hat{H}_0$ and electrostatic potential $V_\mathscr{E}(x)$ in the Hamiltonian $\hat{H}$ from Eq.~\eqref{Hamilton1} are given as
\begin{equation}\label{NormalizedHamiltonian0}
\hat{H}_0=-\frac{1}{\pi^2}\frac{d^2}{dx^2}+V_{QW}(x)
\end{equation}
\begin{equation}\label{NormalizedV1}
V_\mathscr{E}(x)=-\mathscr{E}x
\end{equation}
with
\begin{equation}\label{NormalizedPotentialQW1}
V_{QW}(x)=\left\{
\begin{array}{cc}
0,&|x|<1/2\\
\infty,&|x|\geq1/2.
\end{array}
\right.
\end{equation}
BCs are:

for the Dirichlet QW:
\begin{subequations}\label{BCnormalized1}
\begin{eqnarray}\label{BCnormalized1_DD}
\Psi\!\left(-\frac{1}{2}\right)=\Psi\!\left(\frac{1}{2}\right)=0\\
\textnormal{for the ND case:}\nonumber\\
\label{BCnormalized1_ND}
\Psi'\!\left(-\frac{1}{2}\right)=\Psi\!\left(\frac{1}{2}\right)=0\\
\textnormal{for the DN case:}\nonumber\\
\label{BCnormalized1_DN}
\Psi\!\left(-\frac{1}{2}\right)=\Psi'\!\left(\frac{1}{2}\right)=0\\
\textnormal{for the Neumann QW:}\nonumber\\
\label{BCnormalized1_NN}
\Psi'\!\left(-\frac{1}{2}\right)=\Psi'\!\left(\frac{1}{2}\right)=0
\end{eqnarray}
\end{subequations}
with the prime denoting a derivative of the function with respect to its argument $x$. Solutions of the Schr\"{o}dinger equation form a countably infinite set with the wavefunctions $\Psi_n^{IJ}(\mathscr{E};x)$ for the $n$th level ($n=0,1,2,\ldots$) written as the linear combination of the Airy functions ${\rm Ai}(z)$ and ${\rm Bi}(z)$ \cite{Abramowitz1}:

for the pure Dirichlet BCs:
\begin{subequations}\label{EigenFunctions1}
\begin{eqnarray}\label{EigenFunctions1_DD}
\Psi_n^{DD}\!\!=C_n^{DD}\!\!\left[{\rm Bi}\!\left(z_{n-}^{DD}\right){\rm Ai}\!\left(z_n^{DD}\right)-{\rm Ai}\!\left(z_{n-}^{DD}\right){\rm Bi}\!\left(z_n^{DD}\right)\right]\\
\textnormal{for the ND case:}\nonumber\\
\label{EigenFunctions1_ND}
\Psi_n^{ND}\!\!=C_n^{ND}\!\!\left[{\rm Bi}\!\left(z_{n-}^{ND}\right){\rm Ai}\!\left(z_n^{ND}\right)-{\rm Ai}\!\left(z_{n-}^{ND}\right){\rm Bi}\!\left(z_n^{ND}\right)\right]\\
\textnormal{for the DN case:}\nonumber\\
\label{EigenFunctions1_DN}
\Psi_n^{DN}\!\!=C_n^{DN}\!\!\left[{\rm Bi}'\!\left(z_{n-}^{DN}\right){\rm Ai}\!\left(z_n^{DN}\right)-{\rm Ai}'\!\left(z_{n-}^{DN}\right){\rm Bi}\!\left(z_n^{DN}\right)\right]\\
\textnormal{for the pure Neumann BCs:}\nonumber\\
\label{EigenFunctions1_NN}
\Psi_n^{NN}\!\!=C_n^{NN}\!\!\left[{\rm Bi}'\!\left(z_{n-}^{NN}\right){\rm Ai}\!\left(z_n^{NN}\right)-{\rm Ai}'\!\left(z_{n-}^{NN}\right){\rm Bi}\!\left(z_n^{NN}\right)\right],
\end{eqnarray}
\end{subequations}
where
\begin{eqnarray}\label{CoefficientsX}
z_n^{IJ}\equiv z_n^{IJ}\!\left(\mathscr{E},E_n^{IJ};x\right)=-\!\left(\pi^2\mathscr{E}\right)^{1/3}\!x-\!\!\left(\frac{\pi}{\mathscr{E}}\right)^{2/3}\!\!E_n^{IJ}\\
\label{CoefficientsZ}
z_{n\pm}^{IJ}\equiv z_n^{IJ}\!\left(\mathscr{E},E_n^{IJ};\mp\frac{1}{2}\right)=\pm\frac{1}{2}\!\left(\pi^2\mathscr{E}\right)^{1/3}-\left(\frac{\pi}{\mathscr{E}}\right)^{2/3}\!\!E_n^{IJ}
\end{eqnarray}
with superscripts $I$ and $J$ taking the values of $D$ and/or $N$. Expressions for the normalization constants $C_n$, which are determined from
\begin{equation}\label{Normalization1}
\int_{-1/2}^{1/2}\Psi_n^2(x)dx=1,
\end{equation}
will be discussed below. The form of the functions from Eqs.~\eqref{EigenFunctions1} automatically satisfies the BC at the right edge, $x=1/2$, for all four configurations. Applying the corresponding requirement to the the second confining surface, one gets transcendental equations for finding eigenenergies $E_n$:

for the pure Dirichlet BCs:
\begin{subequations}\label{EigenEnergy1}
\begin{eqnarray}
\label{EigenEnergy1_DD}
&&{\rm Ai}\!\left(z_{n-}^{DD}\right){\rm Bi}\!\left(z_{n+}^{DD}\right)-{\rm Bi}\!\left(z_{n-}^{DD}\right){\rm Ai}\!\left(z_{n+}^{DD}\right)=0\\
&&\textnormal{for the ND case:}\nonumber\\
\label{EigenEnergy1_ND}
&&{\rm Ai}\!\left(z_{n-}^{ND}\right){\rm Bi}'\!\left(z_{n+}^{ND}\right)-{\rm Bi}\!\left(z_{n-}^{ND}\right){\rm Ai}'\!\left(z_{n+}^{ND}\right)=0\\
&&\textnormal{for the DN case:}\nonumber\\
\label{EigenEnergy1_DN}
&&{\rm Ai}'\!\left(z_{n-}^{DN}\right){\rm Bi}\!\left(z_{n+}^{DN}\right)-{\rm Bi}'\!\left(z_{n-}^{DN}\right){\rm Ai}\!\left(z_{n+}^{DN}\right)=0\\
&&\textnormal{for the pure Neumann BCs:}\nonumber\\
\label{EigenEnergy1_NN}
&&{\rm Ai}'\!\left(z_{n-}^{NN}\right){\rm Bi}'\!\left(z_{n+}^{NN}\right)-{\rm Bi}'\!\left(z_{n-}^{NN}\right){\rm Ai}'\!\left(z_{n+}^{NN}\right)=0.
\end{eqnarray}
\end{subequations}
Eqs.~\eqref{EigenEnergy1} represent corresponding energies $E_n^{IJ}(\mathscr{E})$ as implicit functions $F^{IJ}$ of the electric induction $\mathscr{E}$:
\begin{equation}\label{ImplicitF1}
F(E_n,\mathscr{E})=0.
\end{equation}

Calculation of the coefficients $C_n$ is greatly facilitated by the fact that an indefinite integral of the products of the two  linear combinations of the Airy functions ${\rm Ai}(x+\alpha_i)$ and ${\rm Bi}(x+\alpha_i)$, $i=1,2$, with, in general, different constants $\alpha_1$ and $\alpha_2$ is expressed through permutations of the products of the same combinations and their derivatives \cite{Vallee1}. This, together with the  BCs from~\eqref{BCnormalized1} and following from them Eqs.~\eqref{EigenEnergy1} leads to quite simple expressions for $C_n$; for example, for the Dirichlet QW it is:
\begin{equation}\label{Coefficients1_DD}
C_n^{DD}=-\pi\left(\pi^2\mathscr{E}\right)^{1/6}\frac{{\rm Ai}\!\left(z_{n+}^{DD}\right)}{\left[{\rm Ai}^2\!\left(z_{n+}^{DD}\right)-{\rm Ai}^2\!\left(z_{n-}^{DD}\right)\right]^{1/2}},
\end{equation}
where also the value of the Wronskian of the Airy functions, which is equal to $1/\pi$ \cite{Abramowitz1}, has been used. Note that the sign in Eq.~\eqref{Coefficients1_DD} is chosen in such a way that the functions in the vicinity of the right edge do not become negative, $\Psi_n(\mathscr{E};x\rightarrow1/2)\geq0$, what is more convenient for our subsequent discussion.  This completes the calculation of the normalized wavefunctions $\Psi_n(x)$, which correspond  to the energies $E_n$, and ushers in the first method of finding the polarization $P_n$ that in our dimensionless units reads:
\begin{equation}\label{NormalizedPolarization1}
P(\mathscr{E})=\left<x\right>_\mathscr{E}-\left<x\right>_{\mathscr{E}=0}.
\end{equation}
It relies on the direct calculation of the integrals in Eq.~\eqref{NormalizedPolarization1}. First, one needs to find their values at zero electric field. It is important to underline that the well known for this case expressions for the energies
\begin{eqnarray}
E_n^{DD}(0)=(n+1)^2,\,E_n^{DN}(0)=E_n^{ND}(0)=\left(\!n+\frac{1}{2}\right)^2,\nonumber\\
\label{EnergiesZeroField1}
E_n^{NN}(0)=n^2
\end{eqnarray}
and functions
\begin{subequations}\label{FunctionsZeroField1}
\begin{eqnarray}\label{FunctionsZeroField1_DD}
\Psi_n^{DD}(0;x)&=&-2^{1/2}\sin\pi(n+1)\!\left(\!x-\frac{1}{2}\right)\\
\label{FunctionsZeroField1_ND}
\Psi_n^{ND}(0;x)&=&-2^{1/2}\sin\pi\!\left(\!n+\frac{1}{2}\right)\!\left(\!x-\frac{1}{2}\right)\\
\label{FunctionsZeroField1_DN}
\Psi_n^{DN}(0;x)&=&2^{1/2}\cos\pi\!\left(\!n+\frac{1}{2}\right)\!\left(\!x-\frac{1}{2}\right)\\
\label{FunctionsZeroField1_NN}
\Psi_n^{NN}(0;x)&=&\left\{\begin{array}{cc}
1,&n=0\\
2^{1/2}\cos\pi n\!\left(x-\frac{1}{2}\right),&n\geq1\end{array}\right.
\end{eqnarray}
\end{subequations}
can be derived also with the help of the asymptotic properties of the Airy functions \cite{Abramowitz1} as the limit $\mathscr{E}\rightarrow0$ of the corresponding dependencies from Eqs.~\eqref{EigenEnergy1} and \eqref{EigenFunctions1}, respectively. Close enough to the right edge, the functions from Eqs.~\eqref{FunctionsZeroField1} stay positive for any permutation of the BCs and the arbitrary  $n$, $\Psi_n(0;x\rightarrow1/2)\geq0$. Elementary calculation yields:
\begin{subequations}\label{PolarizationsZeroField1}
\begin{eqnarray}\label{PolarizationsZeroField1_Uniform}
\left<x\right>_{\mathscr{E}=0}^{DD}=\left<x\right>_{\mathscr{E}=0}^{NN}&=&0\\
\label{PolarizationsZeroField1_Zaremba}
\left<x\right>_{\mathscr{E}=0}^{DN}=-\left<x\right>_{\mathscr{E}=0}^{ND}&=&\frac{1}{2\pi^2(n+1/2)^2}.
\end{eqnarray}
\end{subequations}
These expected results are naturally explained by the symmetry of the flat QW with respect to its middle for the uniform BCs and the lack of it for the Zaremba configurations. Applied electric field destroys (for the DD and NN structures) this symmetry or changes (for the mixed BCs) the asymmetry pushing in this way the polarization from its zero value. Calculation of the corresponding integrals in Eq.~\eqref{NormalizedPolarization1} can again be performed analytically \cite{Vallee1} producing, after some simple algebra:
%
%\begin{strip}
\begin{subequations}\label{NormalizedPolarization2}
\begin{eqnarray}\label{NormalizedPolarization2DD}
&&P_n^{DD}(\mathscr{E})=-\frac{2}{3}\frac{E_n^{DD}}{\mathscr{E}}
+\frac{1}{6}\frac{{\rm Ai}^2\!\left(z_{n+}^{DD}\right)+{\rm Ai}^2\!\left(z_{n-}^{DD}\right)}{{\rm Ai}^2\!\left(z_{n+}^{DD}\right)-{\rm Ai}^2\!\left(z_{n-}^{DD}\right)}\\
&&P_n^{ND}(\mathscr{E})=-\frac{E_n^{ND}}{\mathscr{E}}+\frac{1}{2\pi^2(n+1/2)^2}\nonumber\\
\label{NormalizedPolarization2ND}
&&-\frac{1}{3}\left(\pi^2\mathscr{E}\right)^{-1/3}\frac{z_{n-}^{ND}+\left[z_{n+}^{ND}{\rm Bi}\!\left(z_{n-}^{ND}\right)\!/{\rm Bi}'\!\left(z_{n+}^{ND}\right)\right]^2}{1+z_{n+}^{ND}\left[{\rm Ai}\!\left(z_{n-}^{ND}\right)\!/{\rm Ai}'\!\left(z_{n+}^{ND}\right)\right]^2}\\
&&P_n^{DN}(\mathscr{E})=-\frac{E_n^{DN}}{\mathscr{E}}-\frac{1}{2\pi^2(n+1/2)^2}\nonumber\\
\label{NormalizedPolarization2DN}
&&-\frac{1}{3}\left(\pi^2\mathscr{E}\right)^{-1/3}\frac{\left(z_{n-}^{DN}\right)^2+z_{n+}^{DN}\left[{\rm Bi}'\!\left(z_{n-}^{DN}\right)\!/{\rm Bi}\!\left(z_{n+}^{DN}\right)\right]^2}{z_{n-}^{DN}+\left[{\rm Ai}'\!\left(z_{n-}^{DN}\right)\!/{\rm Ai}\!\left(z_{n+}^{DN}\right)\right]^2}\\
&&P_n^{NN}(\mathscr{E})=-\frac{E_n^{NN}}{\mathscr{E}}\nonumber\\
\label{NormalizedPolarization2NN}
&&-\frac{1}{3}\left(\pi^2\mathscr{E}\right)^{-1/3}\frac{\left[z_{n+}^{NN}{\rm Bi}'\!\left(z_{n-}^{NN}\right)\right]^2-\left[z_{n-}^{NN}{\rm Bi}'\!\left(z_{n+}^{NN}\right)\right]^2}{z_{n+}^{NN}\left[{\rm Bi}'\!\left(z_{n-}^{NN}\right)\right]^2-z_{n-}^{NN}\left[{\rm Bi}'\!\left(z_{n+}^{NN}\right)\right]^2}.
\end{eqnarray}
\end{subequations}
%\end{strip}
%
Another way of arriving at the same equations will be given in the next section.

Important characteristics of the physical and chemical systems is their quantum information entropy, which can be calculated in the position as well as momentum coordinates. For the 1D QW under consideration, the position $S_x$ and momentum $S_k$ entropies are defined as
\begin{eqnarray}\label{EntropyX_1}
S_x=-\int_{-1/2}^{1/2}\rho(x)\ln\rho(x)dx\\
\label{EntropyK_1}
S_k=-\int_{-\infty}^{\infty}\gamma(k)\ln\gamma(k)dk,
\end{eqnarray}
where $\rho(x)$ and $\gamma(k)$ are, respectively, position and momentum space densities:
\begin{eqnarray}\label{densityX_1}
\rho(x)=|\Psi(x)|^2\\
\label{densityK_1}
\gamma(k)=|\Phi(k)|^2
\end{eqnarray}
with momentum space wave function $\Phi(k)$ being a Fourier transform of its position counterpart:
\begin{equation}\label{Phi1}
\Phi(k)=\frac{1}{\left(2\pi\right)^{1/2}}\int e^{-ikx}\Psi(x)dx.
\end{equation}

Paramount physical significance of these quantities is due to the fact that the total information entropy 
\begin{equation}\label{EntropyTotal1}
S_t=S_x+S_k
\end{equation}
allows one to express the quantum uncertainty relations in the form
\begin{equation}\label{EntropicInequality1}
S_t\ge1+\ln\pi.
\end{equation}
Instructive history of the derivation of the $d$-dimensional analogue of Eq.~\eqref{EntropicInequality1} is described in Ref.~\cite{Bialynicki1}; namely, the first conjectures of these relations from the late 1950-ies \cite{Everett1,Hirschman1} were rigorously confirmed in 1975 by Bia{\l}ynicki-Birula and Mycielski \cite{Bialynicki2} and Beckner \cite{Beckner1}. It is known that the entropic uncertainty relation \cite{Bialynicki3,Wehner1} from Eq.~\eqref{EntropicInequality1} is stronger than its Heisenberg counterpart that is written in the form
\begin{equation}\label{Heisenberg1}
\Delta x\Delta k\ge\frac{1}{2}
\end{equation}
with $\Delta x$ and $\Delta k$ being position and momentum standard deviations:
\begin{eqnarray}\label{DeltaX1}
\Delta x=\sqrt{\left<x^2\right>-\left<x\right>^2}\\
\label{DeltaK1}
\Delta k=\sqrt{\left<k^2\right>-\left<k\right>^2},
\end{eqnarray}
where the momentum averaging is performed on the basis of the corresponding functions $\Phi(k)$:
\begin{equation}\label{Kaveraging}
\left<k^n\right>=\int_{-\infty}^\infty k^n\gamma(k)dk.
\end{equation}
Shortcoming of Eq.~\eqref{Heisenberg1} as compared to the entropy relation \eqref{EntropicInequality1} can be shown for our system as well; consider, for example, the field-free lowest NN state \cite{Bialynicki3,Bialynicki4} that due to the constant unit magnitude position function $\Psi_0^{NN}(x)\equiv1$ exhibits the following momentum density:
\begin{equation}\label{Gamma0NN}
\gamma_0^{NN}(k)=\frac{1}{2\pi}\left(\frac{2}{k}\,\sin\frac{k}{2}\right)^2 \quad{\rm at}\,\,\mathscr{E}=0.
\end{equation}
Obviously,  Heisenberg inequality \eqref{Heisenberg1} in this case is meaningless since the momentum deviation $\Delta k$ diverges due to the infinite integral for $\left<k^2\right>$. However, the corresponding entropy $S_{k_0}^{NN}$ is finite and can be calculated numerically producing the result of $2.6834$, which, together with the zero position entropy $\left. S_{x_0}^{NN}\right|_{\mathscr{E}=0}=0$, does satisfy inequality~\eqref{EntropicInequality1} with its right-hand side equal to $2.1447$. Note that for Eq.~\eqref{Heisenberg1} the situation is aggravated even worse if, instead of taking a momentum expectation value on the basis of the functions $\Phi(k)$ (where $k$ is a number), one employs the averaging on the basis of the position functions $\Psi(x)$ when the momentum is an operator, $\hat{k}\equiv-i\partial_x$:
\begin{equation}\label{KaveragingX}
\left<k\right>_x=\int_{-1/2}^{1/2}\Psi^*(x)\hat{k}\Psi(x)dx.
\end{equation}
For the configuration considered above this immediately leads to $\Delta k=0$ what obviously violates \eqref{Heisenberg1}. It was shown recently \cite{AlHashimi1} that this discrepancy is resolved if one recalls that the Heisenberg inequality \eqref{Heisenberg1} was derived for the infinite volume where the wave function vanishes at infinity. For the non-Dirichlet edge requirements of the finite 1D system a generalized BC-dependent uncertainty condition was derived \cite{AlHashimi1} that for the lowest field-free Neumann state produces $\left.E_0^{NN}\right|_{\mathscr{E}=0}\geq0$, which is indeed satisfied as an equality. For the Dirichlet conditions, the Heisenberg relation remains true since the function $\Psi(x)$ vanishes at the surface and, accordingly, at the infinity; moreover, in this case the momentum deviation $\Delta k$ is the same either calculated over the position $\Psi(x)$ or momentum $\Phi(k)$ functions:
\begin{equation}\label{DeviationDirichletK}
\Delta k_n^{DD}=\pi\left(E_n^{DD}+\mathscr{E}P_n^{DD}\right)^{1/2}.
\end{equation}
\begin{figure}
\centering
\includegraphics[width=0.8\columnwidth]{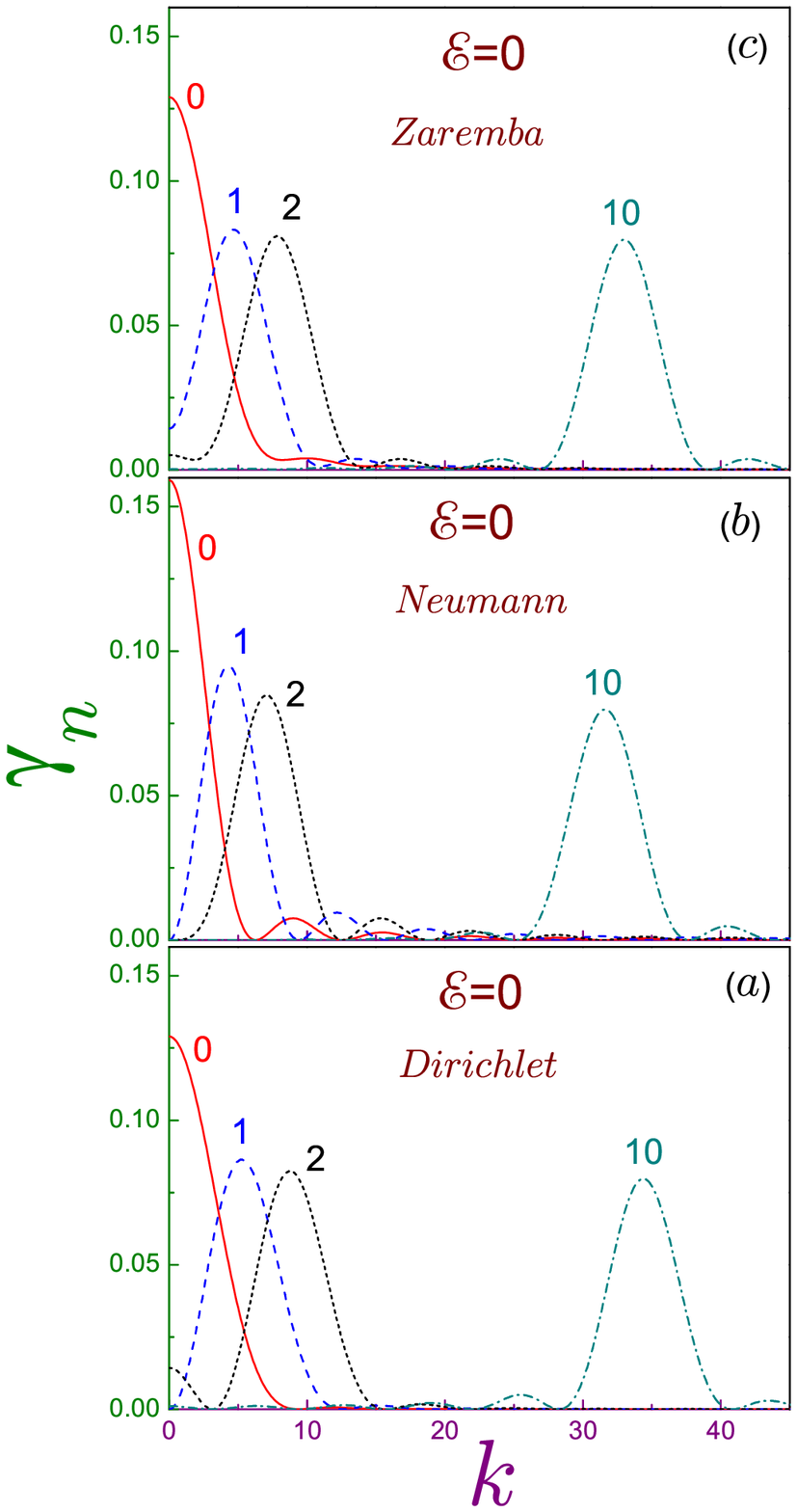}
\caption{\label{MomentumDensitiesZeroField}
Momentum space densities $\color{pakistangreen}\gamma_n$ at zero electric field for (a) the Dirichlet, (b) Neumann, and (c) mixed BCs as functions of the momentum $\color{palatinatepurple}k$ where the solid lines are for the ground states ($n=0$), dashed curves - for the first excited levels ($n=1$), dotted lines - for the second excited states ($n=2$), and dash-dotted curves - for the levels with the quantum number $n=10$.}
\end{figure}

Arguments outlined in the previous paragraph dictate a necessity of studying quantum information entropies, especially for the structures with the non-Dirichlet BCs. Miscellaneous aspects of the properties of the densities $\rho(x)$ and $\gamma(k)$ and entropies $S_x$ and $S_k$ for different quantum systems were addressed in the literature \cite{Gadre1,Robinett1,Tao1,SanchezRuiz1,Majernik1,Majernik2,Majernik3,Dehesa1,LopezRosa1,Aptekarev1,Laguna1}. Before discussing the electric field influence on the densities and entropies, it does make sense to investigate these quantities for the flat QWs. Relevant to the present discussion, let us mention that the position and momentum entropies of the Dirichlet structure were calculated numerically  \cite{Majernik1,Majernik3} and it was shown analytically \cite{SanchezRuiz1} that $S_x^{DD}=\ln2-1\approx-0.30685$, independent of the quantum number $n$. Applying known integrals from the famous reference book \cite{Prudnikov1}, it is easy to show that at the zero fields the same is true for all other BCs but the ground Neumann state, which, of course, is zero:
\begin{equation}\label{EntropyX_ZeroField}
\left.S_x^{IJ}\right|_{\mathscr{E}=0}=\left\{\begin{array}{cc}
0,& n=0,\,I=J=N\\
\ln2-1,&{\rm all\,\,\, other\,\,\, cases.}
\end{array} \right.
\end{equation}
On the other hand, for the momentum entropies at the zero voltage with the densities
\begin{subequations}\label{densityK_2}
\begin{eqnarray}\label{densityK_DD1}
\gamma_n^{DD}(k)&=&\frac{2(n+1)^2\pi\left[1+(-1)^n\cos k\right]}{\left[k^2-(n+1)^2\pi^2\right]^2}\\
\label{densityK_NN1}
\gamma_{n>0}^{NN}(k)&=&\frac{2k^2\left[1-(-1)^n\cos k\right]}{\pi\left(k^2-n^2\pi^2\right)^2}\\
\gamma_n^{ND}(k)&=&\gamma_n^{DN}(k)\nonumber\\
\label{densityK_ND1}
&=&\frac{2(-1)^{n+1}\left(n+\frac{1}{2}\right)\pi k\sin k+k^2+\left(n+\frac{1}{2}\right)^2\pi^2}{\pi\left[k^2-\left(n+\frac{1}{2}\right)^2\pi^2\right]^2}
\end{eqnarray}
\end{subequations}
and for any integrals from Eqs.~\eqref{EntropyX_1}, \eqref{EntropyK_1} and \eqref{Phi1} at $\mathscr{E}\neq0$ there are no analytical expressions in known to us literature \cite{Abramowitz1,Vallee1,Prudnikov1,Prudnikov2_1,Prudnikov3_1,Brychkov1,Gradshteyn1}; accordingly, their direct numerical quadrature was performed in calculating the results presented below. Physically, zero entropy in Eq.~\eqref{EntropyX_ZeroField} means that the uncertainty of determining the position of the particle is the largest one.

Fig.~\ref{MomentumDensitiesZeroField} exhibits zero-field momentum densities $\gamma_n(k)$ from Eqs.~\eqref{densityK_2} and \eqref{Gamma0NN} for all BCs and several quantum states $n$. Previous analysis discussed these dependencies for the Dirichlet QW only \cite{Robinett1,Majernik1,Majernik2,Majernik3}. As $\gamma_n(k)$ is an even function of its argument, $\gamma_n(-k)=\gamma_n(k)$, we plot the parts with the non negative $k$ only. Qualitatively, this behavior is the same for any type of the surface requirements; namely, for all excited states, $n\geq1$, the momentum density is characterized by the two conspicuous peaks, which are located symmetrically with respect to $k=0$  and are flanked by the series of much smaller extrema. The distance between the sharp maxima decreases for the smaller $n$, and for the ground state they merge together forming one symmetric peak with its magnitude being the largest as compared to the excited levels. Quantitatively, Neumann maximum $\gamma_0^{NN}(0)=1/(2\pi)=0.1592$ is larger than the Dirichlet or Zaremba one, $\gamma_0^{DD}(0)=\gamma_0^{ND}(0)=4/\pi^3=0.1290$. As the momentum density is smaller than one, $0\le\gamma_n(k)<1$, the corresponding entropy $S_k$ from Eq.~\eqref{EntropyK_1} is always positive. More detailed numerical analysis reveals that it is a growing function of the quantum number $n$. In contrast, the parts of the position density, which are larger than unity, in their contribution to the integral from Eq.~\eqref{EntropyX_1} overweigh those with $\rho_n(x)\le1$ leading in this way to the negative entropies from Eq.~\eqref{EntropyX_ZeroField}. Returning to the momentum entropies, we note that despite the fact that $\gamma_0^{ND}(0)<\gamma_0^{NN}(0)$, the ground-state entropy $S_{k_0}^{ND}=2.9000$ is greater than its mentioned above Neumann counterpart what apparently is explained by the far-reaching spread of the corresponding density: the wide sidelobe is clearly seen for the $n=0$ curve in panel (c) of Fig.~\ref{MomentumDensitiesZeroField}. In this way, the Dirichlet ground state with its smallest momentum entropy $S_{k_0}^{DD}=2.5189$ comes closest to the limit imposed by the entropic uncertainty relation: $\left. S_{t_0}^{DD}\right|_{\mathscr{E}=0}=2.2120$. Our calculations indicate that the Dirichlet momentum and total entropy for any level $n$ remain the smallest ones among all possible BC combinations: $S_{t_n}^{DD}<S_{t_n}^{_{NN}^{ND}}$.

\begin{figure}
\centering
\includegraphics[width=\columnwidth]{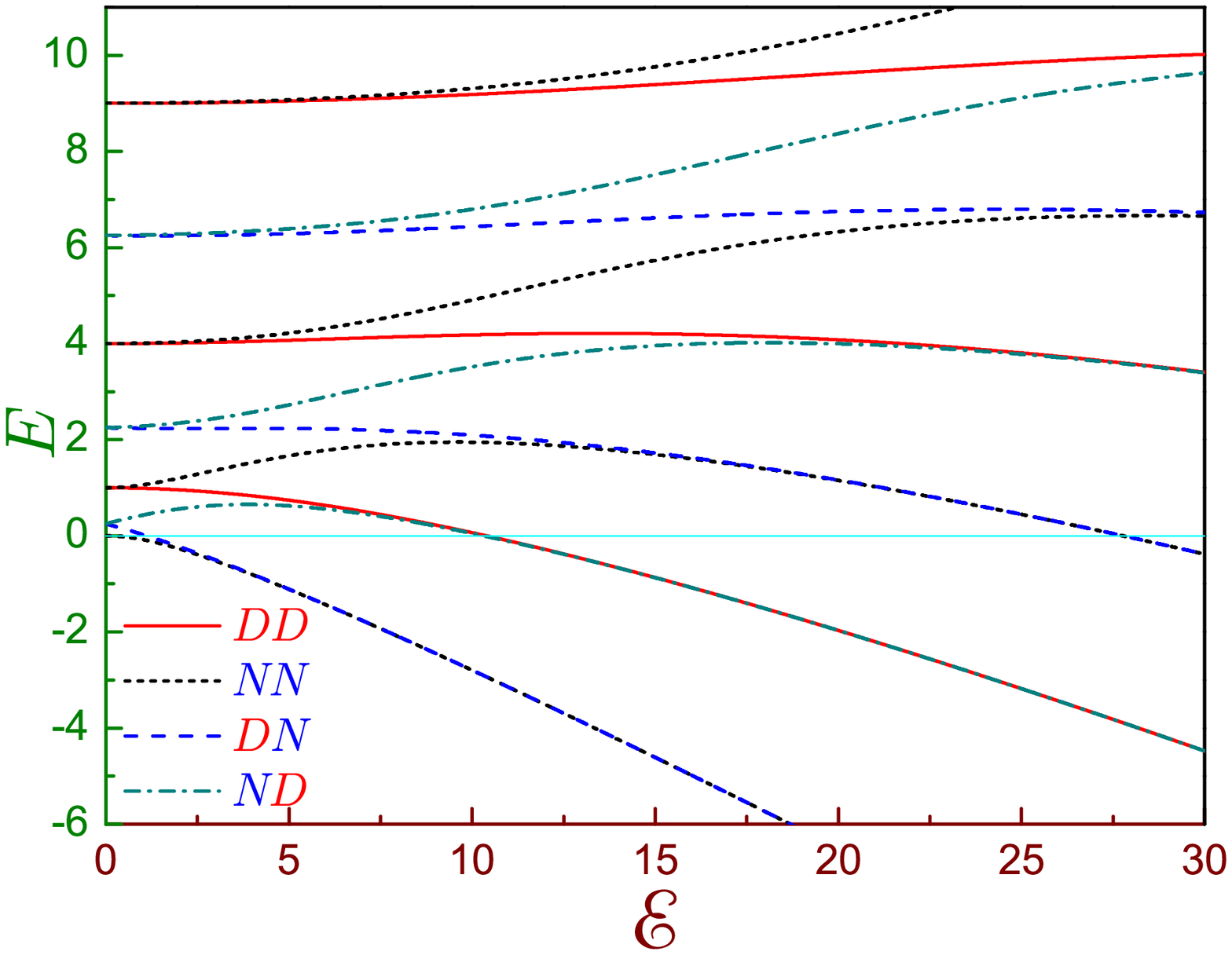}
\caption{\label{Energies}
Energy spectrum $\color{pakistangreen}E$ as a function of the electric field $\color{sangria}\mathscr{E}$ for the pure Dirichlet (solid lines), Neumann (dotted curves), Dirichlet-Neumann (dashed lines) and Neumann-Dirichlet (dash-dotted curves) QW. Horizontal straight line denotes zero energy.}
\end{figure}

\section{Results and discussion}\label{sec_Results}
\begin{figure}
\centering
\includegraphics[width=0.7\columnwidth]{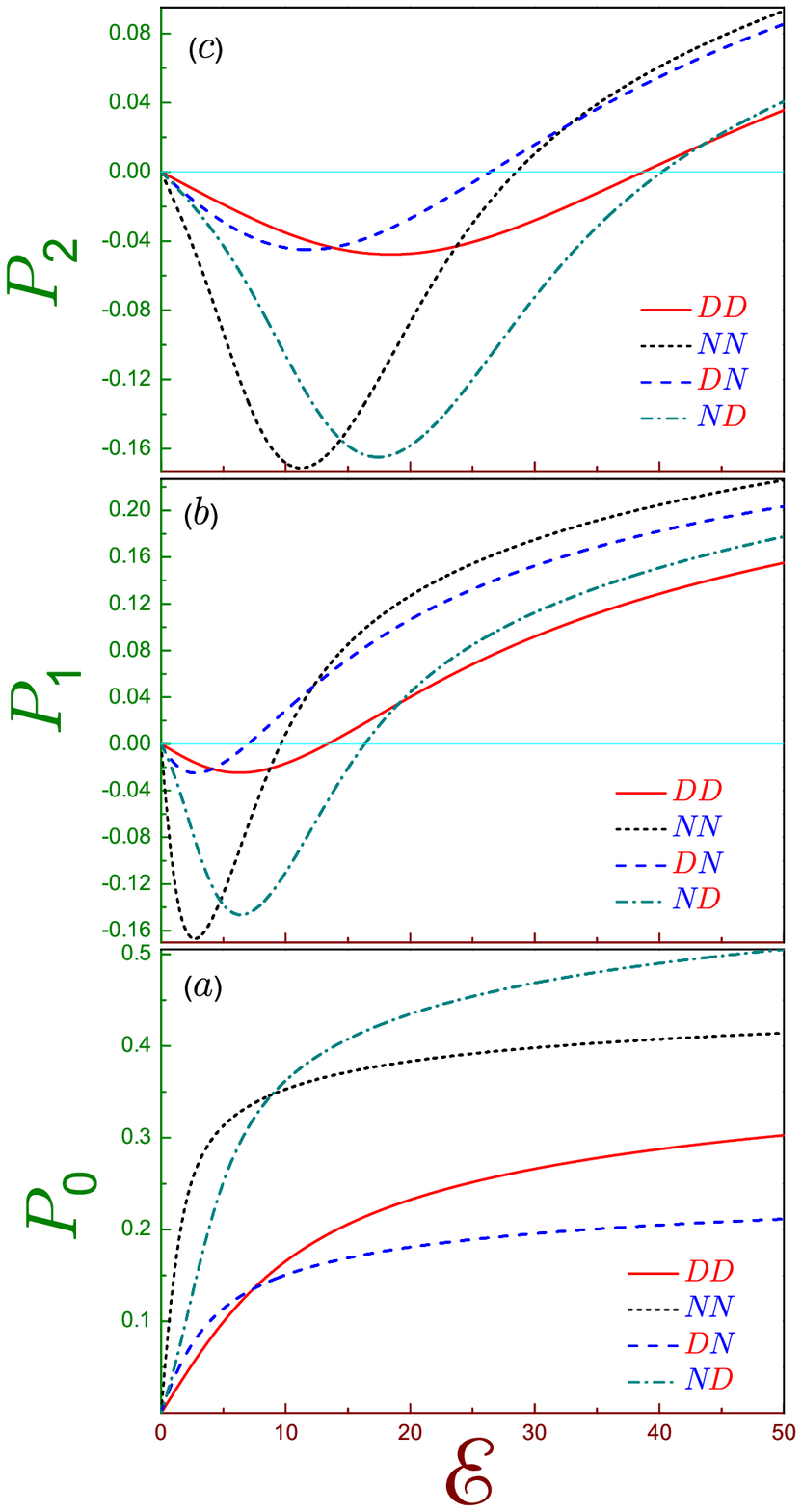}
\caption{\label{Polarizations}
Polarizations $\color{pakistangreen}P_n$ for (a) the ground, $\color{pakistangreen}n=0$, (b) first, $\color{pakistangreen}n=1$, and (c) second, $\color{pakistangreen}n=2$, excited states as functions of the normalized electric field $\color{sangria}\mathscr{E}$. Solid (dotted) lines denote pure Dirichlet (Neumann) configuration while the dashed (dash-dotted) curves are for the DN (ND) geometry. Thin horizontal lines in parts (b) and (c) denote zero polarizations.  Note different $\color{pakistangreen}P$ scales in each of the panels.}
\end{figure}
\subsection{Energy and polarization}
Energy spectrum as a function of the field $\mathscr{E}$ is shown in Fig.~\ref{Energies} for all four types of the possible combinations of the BCs, and corresponding polarizations for the ground and two lowest excited states are depicted in Fig.~\ref{Polarizations}. It is seen that the energies, which at the zero field have the same values for the opposite Zaremba geometries, Eq.~\eqref{EnergiesZeroField1}, at the small voltage move in the opposite directions while at the large $\mathscr{E}$ they coincide with their counterparts for the uniform BC distribution that is imposed at the right surface of the asymmetric edge requirements. This means that the ND geometry is the only configuration at which the lowest energy at the small fields increases. Ground-state polarizations for all BCs monotonically increase with the field and approach at the strong voltages the asymptotic values that will be discussed below while for the excited levels the polarizations take negative values for the small and moderate fields, and only after reaching minimum they increase with the growing $\mathscr{E}$. Classically, this result, which first was predicted for the Dirichlet QW \cite{Nguyen1}, is unexpected since it means that the charged particle moves against the applied force. However, in the quantum mechanical treatment the electron is described as a wave whose behavior is governed by the solution of the corresponding equation inside the domain with the appropriate BCs. For the small electric fields, the energy and polarization dependencies are conveniently described by the standard perturbation theory \cite{Landau1}; namely, considering in the total Hamiltonian $\hat{H}$ from Eq.~\eqref{Hamilton1} the second term as the small disturbance of the field-free Hamiltonian $\hat{H}_0$, one writes the perturbed energies and wavefunctions as
\begin{eqnarray}\label{SmallPerturbation1_Energies}
E_n(\mathscr{E})&=&E_n(0)\!+\! V_{nn}^\mathscr{E}+{{\sum_{m=0}^\infty}}'\frac{\left|V_{mn}^\mathscr{E}\right|^2}{E_n(0)-E_m(0)}\\
\label{SmallPerturbation1_Functions}
\Psi_n(\mathscr{E};x)&=&\Psi_n(0;x)+\!{\sum_{m=0}^\infty}'\frac{V_{mn}^\mathscr{E}}{E_n(0)-E_m(0)}\Psi_m(0;x).
\end{eqnarray}
These expressions are valid at $\left|\frac{V_{mn}^\mathscr{E}}{E_n(0)-E_m(0)}\right|\ll1$. Here, the prime means that the term with $m=n$ is dropped out from the series, and $V_{mn}^\mathscr{E}$ are the matrix elements of the electric potential, Eq.~\eqref{NormalizedV1}, which are calculated on the basis of the undisturbed functions $\Psi_n(0;x)$ from Eqs.~\eqref{FunctionsZeroField1}:
\begin{eqnarray}
V_{mn}^\mathscr{E}=\int_{-1/2}^{1/2}\Psi_m(0;x)V_\mathscr{E}(x)\Psi_n(0;x)dx\nonumber\\
=-\mathscr{E}\int_{-1/2}^{1/2}x\Psi_m(0;x)\Psi_n(0;x)dx.
\end{eqnarray}
Note that the condition of the applicability of the perturbation theory reduces for the low-lying states basically to $\mathscr{E}\ll1$ that, for simplicity, will be used below; however, one has to remember that for the high-lying levels, $n\gg1$, the approximation produces very good results at the $n$-dependent applied voltage $\mathscr{E}$ that actually can be much larger than unity. Integration immediately shows that for the symmetric BCs the energies for the quite small fields are quadratically proportional to $\mathscr{E}$ while for the Zaremba geometries they contain also a linear dependence on the voltage:
\begin{subequations}\label{EnergySmallFields1}
\begin{eqnarray}
&&E_n^{DD}(\mathscr{E}\ll1)=(n+1)^2+\frac{16(n+1)^2}{\pi^4}\nonumber\\
\label{EnergySmallFields1_DD}
&\times&{\sum_{m=0}^\infty}'\!\left[(-1)^{n+m}-1\right]^2\!\!\frac{(m+1)^2}{\left[(n+1)^2-(m+1)^2\right]^5}\mathscr{E}^2\\
&&E_n^{_{DN}^{ND}}(\mathscr{E}\!\ll1)=\left(n+\frac{1}{2}\right)^2\pm\frac{1}{2\pi^2\left(n+1/2\right)^2}\mathscr{E}\nonumber\\
&+&\frac{1}{\pi^4}{\sum_{m=0}^\infty}'\!\!\left\{\frac{\left[(-1)^{n+m}-1\right]^2}{(n-m)^5(n+m+1)}\right.\nonumber\\
\label{EnergySmallFields1_DN}
&+&\left.\!\frac{\left[(-1)^{n+m}+1\right]^2}{(n-m)(n+m+1)^5}\right\}\mathscr{E}^2\\
&&E_n^{NN}(\mathscr{E}\!\ll1)\!=n^2\!+\!\frac{2\!-\!\delta_{n0}}{\pi^4}\nonumber\\
\label{EnergySmallFields1_NN}
&\times&{\sum_{m=0}^\infty}'\!\!(2\!-\!\delta_{m0})\left[(-1)^{n+m}\!-\!1\right]^2\frac{(n^2+m^2)^2}{\left(n^2-m^2\right)^5}\mathscr{E}^2,
\end{eqnarray}
\end{subequations}
where $\delta_{nm}=\left\{\begin{array}{cc}1,&n=m\\0,&n\ne m\end{array}\right.$ is the Kronecker $\delta$. Infinite series in Eqs.~\eqref{EnergySmallFields1} are easily calculated on the basis of the method proposed in Ref.~\cite{Lukes1} for the Dirichlet case. As an example, let us consider its application for the ground state, $n=0$, of the Zaremba geometry. After some simple algebra, the series from~\eqref{EnergySmallFields1_DN} is transformed to
%
%\begin{strip}
\begin{eqnarray}
&-&8\sum_{m=1}^\infty\frac{1}{\left[(2m)^2-1\right]^3}-96\sum_{m=1}^\infty\frac{1}{\left[(2m)^2-1\right]^4}\nonumber\\
\label{{Series1}}
&-&128\sum_{m=1}^\infty\frac{1}{\left[(2m)^2-1\right]^5}.
\end{eqnarray}
%\end{strip}
%
Explicit expression for $\sum_{m=1}^\infty(m^2-a^2)^{-3}$ with the noninteger $a$ is provided in Eq.~5.1.25.37 of Ref. \cite{Prudnikov1_1} with other series being missed there and in other sources; however, for our purposes it is useful to point out that all of them can be readily calculated by observing that each differentiation with respect to $a$ of the identity \cite{Prudnikov2}
\begin{equation}\label{Identity1}
\sum_{m=1}^\infty\frac{1}{m^2-a^2}=\frac{1}{2a^2}-\frac{\pi}{2a}\cot\pi a
\end{equation}
produces one additional power of $m^2-a^2$ in the denominator of the series in its left-hand side. After that, putting $a$ equal to $1/2$ allows to find the expressions above. In the same way, the series $\sum_{m=0}^\infty[(2m+1)^2-a^2]^{-k}$ are found by $k-1$ differentiations of the equality \cite{Prudnikov3}
\begin{equation}\label{Identity2}
\sum_{m=0}^\infty\frac{1}{(2m+1)^2-a^2}=\frac{\pi}{4a}\tan\frac{\pi a}{2}.
\end{equation}
In addition, the series $\sum_{m=0}^\infty(2m+1)^{-6}$ appearing in the calculation of $E_0^{NN}(\mathscr{E})$ can be derived as a limiting case $a\rightarrow0$ of the above mentioned summation, or one can use directly the same source \cite{Prudnikov4} to get its value as $\pi^6 /960$. For reference, the expressions for the three lowest levels at $\mathscr{E}\ll1$ are provided below:
\begin{subequations}\label{EnergySmallFieldsDD}
\begin{eqnarray}\label{EnergySmallFieldsDD0}
E_0^{DD}(\mathscr{E})&=&1-\frac{1}{48}\left(\frac{15}{\pi^2}-1\right)\mathscr{E}^2=1-0.0108\mathscr{E}^2\\
\label{EnergySmallFieldsDD1}
E_1^{DD}(\mathscr{E})&=&4+\left({\frac{1}{192}-\frac{5}{256}\,\frac{1}{\pi^2}}\right)\mathscr{E}^2=4+0.00323\mathscr{E}^2\\
\label{EnergySmallFieldsDD2}
E_2^{DD}(\mathscr{E})&=&9+\frac{1}{2^{\,4}\!\cdot3^3}\left(1-\frac{5}{3}\,\frac{1}{\pi^2}\right)\mathscr{E}^2=9+0.00192\mathscr{E}^2
\end{eqnarray}
\end{subequations}
(Eq.~\eqref{EnergySmallFieldsDD0} was derived before either with the help of the perturbation calculations \cite{Lukes1,Bastard1} or variational approach \cite{Bastard1,Ahn1} or hypervirial-perturbational treatment \cite{Fernandez1})
\begin{subequations}\label{EnergySmallFieldsND}
\begin{eqnarray}
\! E_0^{ND}(\mathscr{E})\! &=& \!\frac{1}{4}\!+\!\frac{2}{\pi^2}\,\mathscr{E}\!+\!\!\left(\frac{1}{12}\!+\!\frac{1}{\pi^2}\!-\!\frac{20}{\pi^4}\right)\!\mathscr{E}^2\nonumber\\
\label{EnergySmallFieldsND0}
&=&\!0.25\!+\!0.203\mathscr{E}\!-\! 0.0207\mathscr{E}^2\\
E_1^{ND}(\mathscr{E})&=&\frac{9}{4}+\frac{2}{9\pi^2}\,\mathscr{E}+\left(\frac{293}{31104}+\frac{1127}{93312}\,\frac{1}{\pi^2}-\frac{20}{729}\,\frac{1}{\pi^4}\right)\mathscr{E}^2\nonumber\\
\label{EnergySmallFieldsND1}
&=&2.25\!+\!0.0225\mathscr{E}\!+\!0.0104\mathscr{E}^2\\
E_2^{ND}(\mathscr{E})&=&\frac{25}{4}+\frac{2}{25\,\pi^2}\,\mathscr{E}+\left(\frac{1}{300}+\frac{1}{625}\,\frac{1}{\pi^2}-\frac{8}{3125}\,\frac{1}{\pi^4}\right)\mathscr{E}^2\nonumber\\
\label{EnergySmallFieldsND2}
&=&6.25+0.00811\mathscr{E}+0.00348\mathscr{E}^2
\end{eqnarray}
\end{subequations}
(expressions for the DN case are given by Eqs.~\eqref{EnergySmallFieldsND} with, however, the negative sign of the linear term)
\begin{subequations}\label{EnergySmallFieldsNN}
\begin{eqnarray}\label{EnergySmallFieldsNN0}
E_0^{NN}(\mathscr{E})&=&-\frac{\pi^2}{120}\,\mathscr{E}^2=-0.0822\mathscr{E}^2\\
\label{EnergySmallFieldsNN1}
E_1^{NN}(\mathscr{E})&=&1+\frac{1}{16}\left(\frac{7}{\pi^2}+\frac{1}{3}\right)\mathscr{E}^2=1+0.0652\mathscr{E}^2\\
\label{EnergySmallFieldsNN2}
E_2^{NN}(\mathscr{E})&=&4+\frac{1}{64}\left(\frac{7}{4\pi^2}+\frac{1}{3}\right)\mathscr{E}^2=4+0.00798\mathscr{E}^2.
\end{eqnarray}
\end{subequations}
Note that in this regime the difference $E_1-E_0$ decreases for the ND geometry and grows with $\mathscr{E}$ for all other BCs what causes different voltage dependence of the critical temperatures of the Bose-Einstein systems in the corresponding 1D box \cite{Olendski2}. Invoking regular units, one can say that, for the energy separation between the two lowest states, the small electrostatic field leads to the decrease of the effective width for all but ND wells.

For calculating the polarizations $P_n$, one can insert the expressions for the wavefunctions from~\eqref{SmallPerturbation1_Functions} into the general definition, Eq.~\eqref{NormalizedPolarization1}, and perform the integration with the subsequent evaluation of the series as it just was described above. However, there is another applicable at the arbitrary fields elegant method of computing them that employs the Hellmann-Feynman theorem \cite{Feynman1,Lowdin1} stating:
\begin{equation}\label{HellmannFeynman1}
\frac{dE}{d\lambda}=\left\langle\Psi_\lambda\left|\frac{\partial\hat{H}_\lambda}{\partial\lambda}\right|\Psi_\lambda\right\rangle.
\end{equation}
In this equation, $\lambda$ is some continuous parameter of the system. Applying it to our configuration where the role of the variable $\lambda$ is played by the electric field one arrives at
\begin{equation}\label{HellmannFeynman2}
\frac{dE_n}{d\mathscr{E}}=\left\langle\frac{\partial\hat{H}}{\partial\mathscr{E}}\right\rangle=-\left\langle x\right\rangle,
\end{equation}
where the last equality follows straightforwardly from Eqs.~\eqref{Hamilton1}, \eqref{NormalizedHamiltonian0} and \eqref{NormalizedV1}. In this way, the following expression for the polarization, which is equivalent to that given above, Eq.~\eqref{NormalizedPolarization1}, is derived:
\begin{equation}\label{NormalizedPolarization3}
P_n(\mathscr{E})=-\frac{dE_n}{d\mathscr{E}}-\left<x\right>_{\mathscr{E}=0}.
\end{equation}
This fundamental relation \cite{Montgomery1} establishes an intimate connection between the physical observable quantity (in our case, polarization) and the speed of the energy change (for our geometry, with the electric field). The magnitude of the energy itself is a relative value that depends on the choice of the origin from which we measure it; for example, for the pure Dirichlet ground state it is a continuously increasing function of the field if zero energy coincides with the  lowest point located at the right edge \cite{Nguyen1}, see Fig.~\ref{Profile}. However, the difference between the two energies has an absolute meaning, as Eq.~\eqref{NormalizedPolarization3} vividly manifests. Applying it to Eqs.~\eqref{EnergySmallFields1}, one gets:
\begin{subequations}\label{PolarizationSmallFields1}
\begin{eqnarray}
P_n^{DD}(\mathscr{E})&=&-\!\frac{32(n+1)^2}{\pi^4}\nonumber\\
\label{PolarizationSmallFields1_DD}
&\times &{\sum_{m=0}^\infty}'\!\!\left[(-1)^{n+m}\!-\!1\right]^2\!\!\frac{(m+1)^2}{\left[(n\!+\!1)^2\!-\!(m\!+\!1)^2\right]^5}\mathscr{E}\\
P_n^{ND}(\mathscr{E})&=&P_n^{DN}(\mathscr{E})=-\frac{2}{\pi^4}\!\!\!\!{\sum_{m=0}^\infty}'\!\!\left\{\frac{\left[(-1)^{n+m}-1\right]^2}{(n-m)^5(n+m+1)}\right.\nonumber\\
\label{PolarizationSmallFields1_ND_DN}
&+&\left.\frac{\left[(-1)^{n+m}+1\right]^2}{(n-m)(n+m+1)^5}\right\}\mathscr{E}\\
P_n^{NN}(\mathscr{E})&=&-\frac{2\left(2\!-\!\delta_{n0}\right)}{\pi^4}\nonumber\\
\label{PolarizationSmallFields1_NN}
&\times &{\sum_{m=0}^\infty}'\!\!\!\!\!\left(2\!-\!\delta_{m0}\right)\left[(-1)^{n+m}-\!1\right]^2\frac{\left(n^2+m^2\right)^2}{\left(n^2-m^2\right)^5}\mathscr{E},
\end{eqnarray}
\end{subequations}
valid at $\mathscr{E}\ll1$. It is important to note that for the uniform BCs the energies at the small fields depend quadratically on the voltage while the polarizations are linear functions of $\mathscr{E}$. This is explained by the symmetry of the QW with respect to its middle, which causes that the first-order energy perturbation $V_{nn}^\mathscr{E}$ vanishes identically. However, the non-diagonal matrix elements of  $V_\mathscr{E}(x)$, which enter into the expression for the perturbed function, Eq.~\eqref{SmallPerturbation1_Functions}, determine that the latter is deformed and, accordingly, the polarization survives already in the first order of the expansion with respect to the small parameter $\mathscr{E}$. For the Zaremba QWs, there is no such symmetry at the zero field; accordingly, both energies as well as polarizations are linear functions of the voltage. It is also notable from Eq.~\eqref{PolarizationSmallFields1_ND_DN} that in this regime the polarizations of the opposite mixed BCs are equal to each other, as it is also seen from Fig.~\ref{Polarizations} at the small fields where they are flanked by the uniform edge requirements. In addition, it is well known that the second-order corrections to the ground-state energy are always negative \cite{Landau1}; as a result, the corresponding polarizations that are defined from Eq.~\eqref{NormalizedPolarization3} increase. 

\begin{figure*}
\centering
\includegraphics[width=\textwidth]{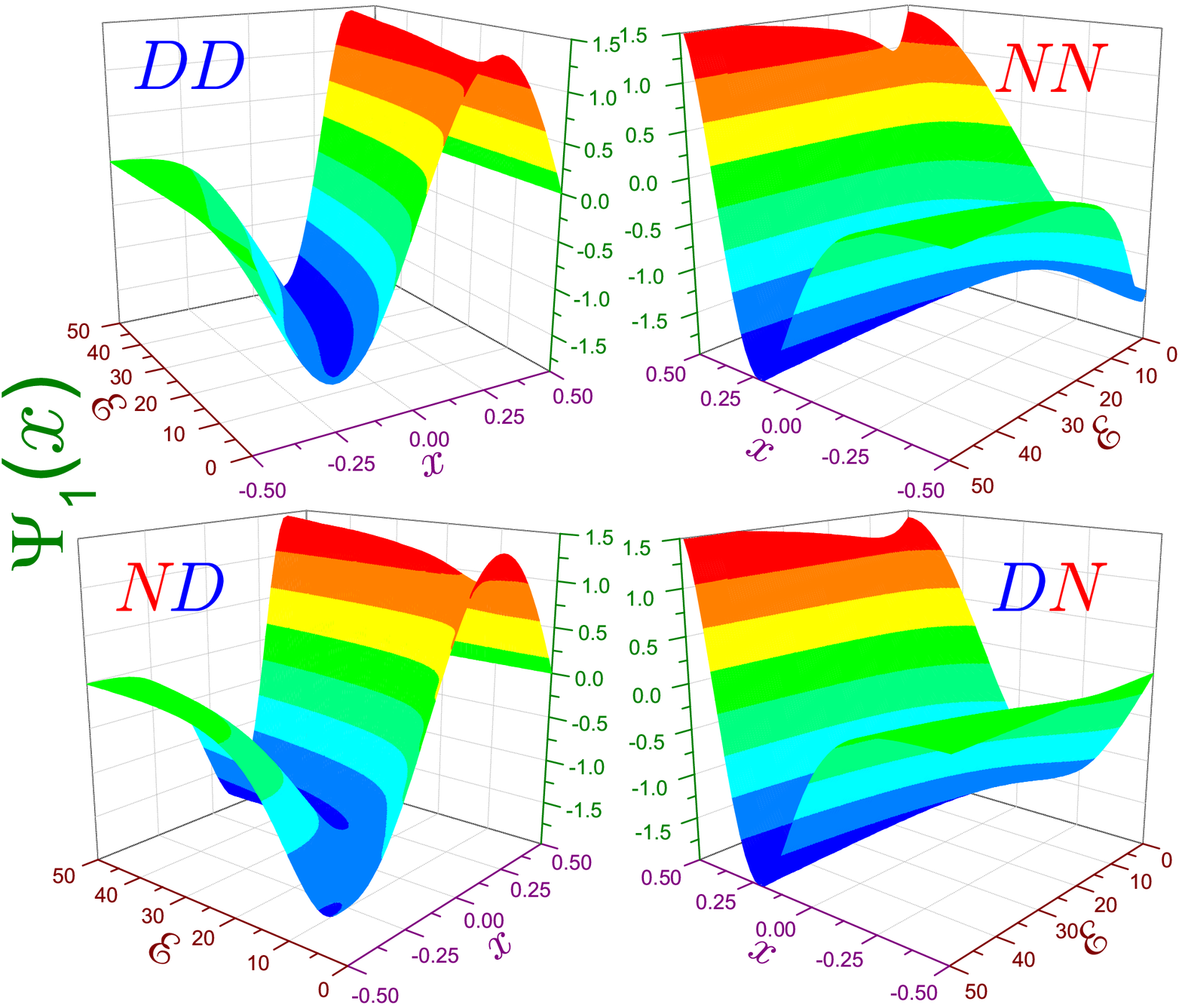}
\caption{\label{Function1}
Function $\color{pakistangreen}\Psi_1({\color{sangria}\mathscr{E}};{\color{palatinatepurple}x})$  of the first excited level in terms of the distance $\color{palatinatepurple}x$ and electric field $\color{sangria}\mathscr{E}$ for all possible permutations of the BCs. In each of the panels, the corresponding type of the edge requirements is denoted by the two characters. To exhibit better the characteristic features of every BC configuration, different viewing perspectives are used in the left and right parts of the figure.}
\end{figure*}

\begin{figure*}
\centering
\includegraphics[width=\textwidth]{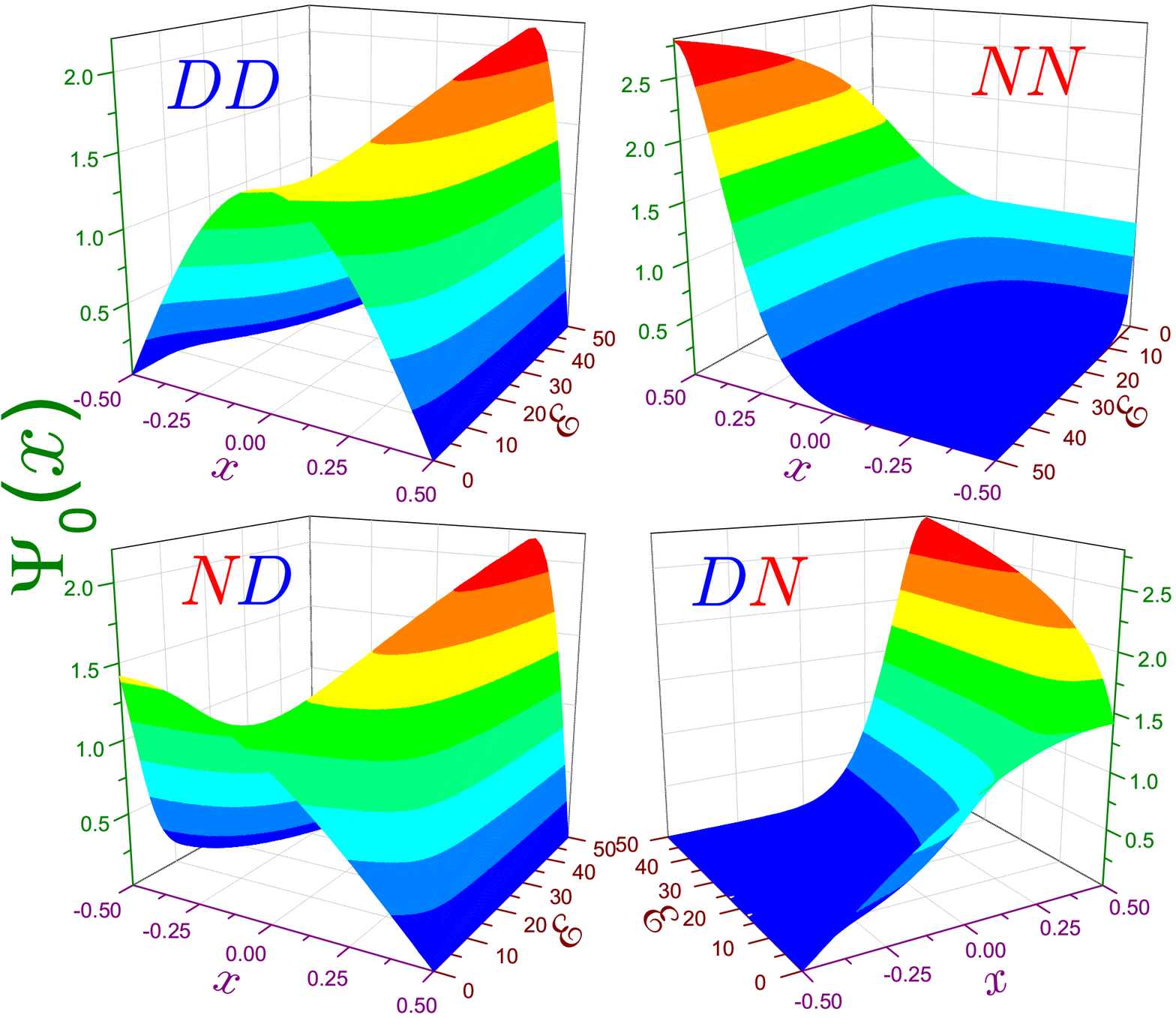}
\caption{\label{Function0}
The same as in Fig.~\ref{Function1} but for the ground state function $\color{pakistangreen}\Psi_0({\color{sangria}\mathscr{E}};{\color{palatinatepurple}x})$. For the DN geometry, the viewing location is different as compared to other BCs.}
\end{figure*}

The easiest way to explain qualitatively (and to the large degree of precision, quantitatively) the negative value of the excited-state polarization at the small and moderate electric fields  lies in the analysis of the first excited level for the pure Neumann BCs. In general, the applied voltage mixes all zero-field states, as Eq.~\eqref{SmallPerturbation1_Functions} shows. But for our consideration it suffices to take into account the field-induced admixture to $\Psi_1$ produced by the ground level since it is the nearest lying state that, accordingly, exerts the largest influence:
\begin{eqnarray}
\Psi_1^{NN}(\mathscr{E};x)&\approx&\Psi_1^{NN}(0;x)+V_{01}^\mathscr{E}\Psi_0^{NN}(0;x)\nonumber\\
\label{ApproximateFunction1}
&=&2^{1/2}\!\left[\cos\pi\!\!\left(\!x\!-\!\frac{1}{2}\right)\!-\!\frac{2}{\pi^2}\mathscr{E}\right],\quad\mathscr{E}\ll1.
\end{eqnarray}
This equation manifests that the small electric admixture of the ground state subtracts the $x$-independent term from $\Psi_1^{NN}(0;x)$. If one recalls that, by our convention, this function is nonnegative at the right interface, $x=1/2$, it means that the applied voltage decreases (increases) the amplitude of the right (left) extremum. As a result, the square of the wavefunction $|\Psi_1^{NN}(\mathscr{E};x)|^2$ gets  smaller (bigger) at the right (left) edge of the QW what classically means the paradoxical shift of the particle against the applied force. However, quantum mechanical treatment considers the electron as a wave and the negative polarization is a manifestation of its wave nature equivalent to the  interference of the modes in the Fabry-P\'{e}rot resonator \cite{Lipson1} with gradually changing inside it index of refraction. The decrease of the NN function at the right wall for the small fields is clearly seen in the corresponding panel of Fig.~\ref{Function1} that shows $\Psi_1(x)$ vs. the electric field for all four distributions of the BCs. Polarization of the state elementary follows from its definition, Eq.~\eqref{NormalizedPolarization1}, and  approximate expression for $\Psi_1^{NN}$ from~\eqref{ApproximateFunction1}:
\begin{equation}\label{ApproximatePolarization1}
P_1^{NN}(\mathscr{E})\approx-\frac{16}{\pi^4}\mathscr{E}=-0.164\mathscr{E},\quad\mathscr{E}\ll1.
\end{equation}
Taking into account all higher lying states modifies insignificantly the above result, as an application of the Hellmann-Feyman theorem, Eq.~\eqref{NormalizedPolarization3}, to \eqref{EnergySmallFieldsNN1} reveals:
\begin{equation}\label{ApproximatePolarization1_1}
P_1^{NN}(\mathscr{E})=-\frac{1}{8}\left(\frac{7}{\pi^2}+\frac{1}{3}\right)\mathscr{E}=-0.130\mathscr{E},\quad\mathscr{E}\ll1.
\end{equation}
Note that the same two-state approximation applied to the ground level produces the following expressions for the function $\Psi_0^{NN}(x)$ and polarization $P_0^{NN}$:
\begin{eqnarray}
\Psi_0^{NN}(\mathscr{E};x)&\approx&\Psi_0^{NN}(0;x)+V_{10}^\mathscr{E}\Psi_1^{NN}(0;x)\nonumber\\
\label{ApproximateFunction0}
&=&1+\frac{4}{\pi^2}\mathscr{E}\cos\pi\!\left(x-\frac{1}{2}\right),\quad\mathscr{E}\ll1
\end{eqnarray}
\begin{equation}\label{ApproximatePolarization0}
P_0^{NN}(\mathscr{E})\approx\frac{16}{\pi^4}\mathscr{E}=0.1642557\mathscr{E},\quad\mathscr{E}\ll1.
\end{equation}
We retained in the last result more digits than before in order to show its excellent coincidence with the one obtained from Eq.~\eqref{EnergySmallFieldsNN0}, which counts the contributions from all states:
\begin{equation}\label{ApproximatePolarization0_0}
P_0^{NN}(\mathscr{E})=\frac{\pi^2}{60}\mathscr{E}=0.1644934\mathscr{E},\quad\mathscr{E}\ll1.
\end{equation}
Comparison of Eqs.~\eqref{ApproximatePolarization0} and \eqref{ApproximatePolarization0_0} proves that for the ground level the dominant admixture comes from the nearest lying excited state while the contributions from the higher levels can be safely neglected. Eq.~\eqref{ApproximateFunction0} tells us that for the lowest level the application of the field leads to such deformation of the particle concentration inside the well that is consistent with the rules of the classical mechanics: probability of finding the electron at the right (left) wall increases (decreases) with the  electric force. Dependence on the field of the ground state wavefunctions $\Psi_0^{IJ}(x)$ for all possible $I$ and $J$ is shown in Fig.~\ref{Function0}.

Returning to the interaction of the right Neumann wall and the field, let us note that each of them separately causes the particle to accommodate closer to the interface: the surface attracts the electron by creating an extremum of its wavefunction while the appropriately directed electric force simply pushes it closer to the border. However, for any excited state their combined influence on the particle in some range of the small voltages leads to the opposite result, i.e., to its repulsion from the wall. We underline again that this is a manifestation of the wave nature of the quantum charge carriers. Fig.~\ref{Function1} shows that for all other combinations of the BCs the closest to the right wall extremum undergoes the same suppression of its magnitude what results in the negative polarizations exhibited by panels (b) and (c) of Fig.~\ref{Polarizations}. Physically, their explanation is absolutely identical to the one discussed above for the purely Neumann QW; namely, it is the field-induced admixture of the different states that causes their destructive interference. Above, we provided the detailed analysis of the NN structure only since it is  the simplest from the mathematical point of view, transparent and very instructive one. It is also clear why at the small fields the magnitude of the polarization for the ground or excited states is the largest for the purely Neumann QW, as Fig.~\ref{Polarizations} depicts; namely, for the flat structure all extrema have the same amplitude equal to $2^{1/2}$, see Eqs.~\eqref{FunctionsZeroField1}, and their relative weight in the integral from \eqref{QMexpectation1} is determined by its distance from the centre of the well: the further from the origin, the larger its contribution is. Obviously, the field-induced change of the maxima or minima located at the Neumann walls will impact the polarization stronger than their Dirichlet counterparts, which are repelled from the boundary into the interior of the well. For the same reason, Zaremba geometries with one attractive and one repulsive surfaces have their polarizations for $\mathscr{E}\ll1$ lying in between those two for the uniform BCs.

Further growth of the electric field makes the perturbation theory less and less applicable, and Eqs.~\eqref{EigenEnergy1} and \eqref{NormalizedPolarization2} should be used for finding energies and polarizations. It is noteworthy to show that the latter dependencies are derivable from the former ones with the help of the Hellmann-Feynman theorem that holds for the arbitrary $\mathscr{E}$. Namely, for finding a derivative that enters \eqref{NormalizedPolarization3} one considers eigenvalue Eqs.~\eqref{EigenEnergy1} as implicit functions of the energy $E(\mathscr{E})$ on the electric field with its most general form provided by Eq.~\eqref{ImplicitF1}. A theorem of differentiating implicit functions states \cite{Rudin1}:
\begin{equation}\label{ImplicitDerivative1}
\frac{dE}{d\mathscr{E}}=-\frac{\partial F/\partial\mathscr{E}}{\partial F/\partial E}.
\end{equation}
Applying this rule to each of Eqs.~\eqref{EigenEnergy1}, one arrives at the corresponding dependence from \eqref{NormalizedPolarization2}. Thus, there are two equivalent methods of calculating the polarization: one implements a direct integration of \eqref{NormalizedPolarization3} with the use of the analytical properties of the primitives of the Airy functions, and the second one relies on the application of the Hellmann-Feynman theorem and the rule for finding a derivative of the implicit functions.

At the very strong fields, it is reasonable to assume that the electron is so squeezed to the right surface that it does not 'feel' the left boundary what leads to the problem of the triangular potential with the vertical Dirichlet \cite{Katriel1} or Neumann right wall.  Then, the wavefunction degenerates to
\begin{equation}\label{FunctionHighFields1}
\Psi_n(\mathscr{E},x)=C_n{\rm Ai}\!\left(-\left(\pi^2\mathscr{E}\right)^{1/3}\!x-\left(\frac{\pi}{\mathscr{E}}\right)^{2/3}\!\!E_n\right),\quad\mathscr{E}\gg1.
\end{equation}
Application of the right BC leads in this limit to the following energies:
\begin{subequations}\label{EnergyHighFields1}
\begin{eqnarray}\label{EnergyHighFields1_D}
E_n^{DD}(\mathscr{E})=E_n^{ND}(\mathscr{E})=-\frac{1}{2}\mathscr{E}-\left(\frac{\mathscr{E}}{\pi}\right)^{2/3}\!a_n,\quad\mathscr{E}\gg1\\
\label{EnergyHighFields1_N}
E_n^{NN}(\mathscr{E})=E_n^{DN}(\mathscr{E})=-\frac{1}{2}\mathscr{E}-\left(\frac{\mathscr{E}}{\pi}\right)^{2/3}a_n^\prime,\quad\mathscr{E}\gg1,
\end{eqnarray}
\end{subequations}
what results in the polarizations
\begin{subequations}\label{PolarizationHighFields1}
\begin{eqnarray}\label{PolarizationHighFields1_DD}
P_n^{DD}(\mathscr{E})&=&\frac{1}{2}+\frac{2}{3}\frac{a_n}{\pi^{2/3}}\mathscr{E}^{-1/3},\quad\mathscr{E}\gg1\\
\label{PolarizationHighFields1_ND}
P_n^{ND}(\mathscr{E})&=&\frac{1}{2}+\frac{2}{3}\frac{a_n}{\pi^{2/3}}\mathscr{E}^{-1/3}+\frac{1}{2\pi^2(n+1/2)^2},\,\mathscr{E}\gg1\\
\label{PolarizationHighFields1_DN}
P_n^{DN}(\mathscr{E})&=&\frac{1}{2}+\frac{2}{3}\frac{a_n^\prime}{\pi^{2/3}}\mathscr{E}^{-1/3}-\frac{1}{2\pi^2(n+1/2)^2},\,\mathscr{E}\gg1\\
\label{PolarizationHighFields1_NN}
P_n^{NN}(\mathscr{E})&=&\frac{1}{2}+\frac{2}{3}\frac{a_n^\prime}{\pi^{2/3}}\mathscr{E}^{-1/3},\quad\mathscr{E}\gg1.
\end{eqnarray}
\end{subequations}
Here, $a_n$ and $a_n^\prime$ are the $n$th solutions of equations \cite{Abramowitz1}
\begin{subequations}
\begin{eqnarray}
{\rm Ai}(a_n)&=&0\\
{\rm Ai}^\prime(a_n^\prime)&=&0,
\end{eqnarray}
\end{subequations}
respectively. Their values are provided, for example, in Table 10.13 of Ref.~\cite{Abramowitz1}. Also, function {\bf AiryAiZeros} of computer algebra system Maple\texttrademark, Ref.~\cite{Maple1}, calculates $a_n$ with any desired precision. Numerically, it is noteworthy to remark that the absolute value of the difference between the two consecutive zeros is smaller for the Airy function in comparison with its derivative:
\begin{equation}\label{difference1}
a_n-a_{n+1}<a_n'-a_{n+1}',
\end{equation}
what leads to the different critical temperatures of the corresponding Bose-Einstein condensates \cite{Olendski2}. Eqs.~\eqref{PolarizationHighFields1} show that there are saturation values of the polarizations in the asymptotics of the very high fields:
\begin{subequations}\label{PolarizationHighFields2}
\begin{eqnarray}\label{PolarizationHighFields2_D}
P_n^{DD}(\infty)&=&P_n^{NN}(\infty)=\frac{1}{2}\\
\label{PolarizationHighFields2_ND}
P_n^{_{DN}^{ND}}(\infty)&=&\frac{1}{2}\pm\frac{1}{2\pi^2(n+1/2)^2}.
\end{eqnarray}
\end{subequations}
Due to small power of $\mathscr{E}$ in \eqref{PolarizationHighFields1}, this limit is approached quite slowly, as Fig.~\ref{Polarizations} shows. On the contrary, the equal values of the energies for the QWs with the same BC at the right surface, Eq.~\eqref{EnergyHighFields1}, are achieved at relatively moderate voltages, which are level dependent: for the higher lying states this happens at the larger fields, see Fig.~\ref{Energies}.

Next, let us discuss the electric fields and corresponding energies at which the polarizations turn to zero. In general, they should be found from the requirement of the vanishing right-hand-side of \eqref{NormalizedPolarization2}; for example, for the purely Dirichlet case, Eq.~\eqref{NormalizedPolarization2DD} yields the following transcendental equation for calculating them:
\begin{equation}\label{ZeroPolarization1}
{\rm Ai}^2\!\left(z_{n-}^{DD}\right)=\frac{4\frac{E_n^{DD}}{\mathscr{E}^{ext}}-1}{4\frac{E_n^{DD}}{\mathscr{E}^{ext}}+1}{\rm Ai}^2\!\left(z_{n+}^{DD}\right),
\end{equation}
where the superscript at $\mathscr{E}$ means that the energy at this electric field reaches maximum. As a first approximation, the high-field limit from Eq.~\eqref{PolarizationHighFields1_DD} provides the value of this voltage as
\begin{equation}\label{PolarizationExtremum_Field1}
\mathscr{E}_n^{ext}\approx\left(\frac{4}{3}\right)^3\frac{|a_n|^3}{\pi^2}
\end{equation}
while the corresponding maximum energy at this point is
\begin{equation}\label{PolarizationExtremum_Energy1}
E_n^{ext}\approx\frac{1}{3}\left(\frac{4}{3}\right)^2\frac{|a_n|^3}{\pi^2},
\end{equation}
where, for convenience, we have substituted the superscript denoting a type of the BC by the one denoting that this is an extremum value. One gets from Eq.~\eqref{PolarizationExtremum_Field1}: $\mathscr{E}_1^{ext}\approx16.407$ and $\mathscr{E}_2^{ext}\approx40.408$ while the exact ones are: $13.370$ and $38.700$. The corresponding approximate energies read: $E_1^{ext}\approx4.102$ and $E_2^{ext}\approx10.102$ with the exact ones being $4.212$ and $10.146$. So, the precision of the approximation increases with the level number and is better for the energy than for the field. It directly follows from Eqs. \eqref{PolarizationExtremum_Field1} and \eqref{PolarizationExtremum_Energy1} that
\begin{equation}\label{Expression1}
4\frac{E_n^{ext}}{\mathscr{E}_n^{ext}}-1=0,
\end{equation}
what automatically zeroes the right-hand side of Eq.~\eqref{ZeroPolarization1}, as it should be, since it contains a contribution from the left wall that is neglected in our high-field approach developed above. Assuming that the expression from Eq.~\eqref{Expression1} is a small nonzero number,
\begin{equation}\label{Expression2}
4\frac{E_n^{ext}}{\mathscr{E}_n^{ext}}-1=\delta,\quad|\delta|\ll1,
\end{equation}
and applying the corresponding Taylor expansion of the left-hand side of Eq.~\eqref{ZeroPolarization1} with respect to the small parameter $\delta$, one can improve the approximations from Eqs.~\eqref{PolarizationExtremum_Field1} and \eqref{PolarizationExtremum_Energy1}; however, the obtained results are not very transparent and, accordingly, not written here.

As a next example, it is instructive to introduce zero-energy electric field $\mathscr{E}_n^{(0)}$, i.e., the field at which the energy of the $n$th level turns to zero. Its physical meaning is obvious from its definition and the transcendental equation for its determination follows from Eqs.~\eqref{EigenEnergy1} by putting $E_n=0$. Assuming that this happens at quite large fields, one can invoke asymptotic properties of the Airy functions \cite{Abramowitz1} to get
\begin{subequations}
\begin{eqnarray}\label{FieldZeroEnergyD}
\left.\mathscr{E}_n^{(0)}\right|_{DD}=\left.\mathscr{E}_n^{(0)}\right|_{ND}\approx18\left(n+\frac{3}{4}\right)^{\!2}\\
\label{FieldZeroEnergyN}
\left.\mathscr{E}_{n+1}^{(0)}\right|_{NN}=\left.\mathscr{E}_n^{(0)}\right|_{DN}\approx18\left(n+\frac{1}{4}\right)^{\!2}
\end{eqnarray}
\end{subequations}
with the obvious requirement $\left.\mathscr{E}_0^{(0)}\right|_{NN}=0$. Even though these dependencies are derived for the high voltages, they produce reasonably good results even for the small $n$; for example, the lowest DN level reaches zero at $\mathscr{E}_0^{(0)}=1.118$ while Eq.~\eqref{FieldZeroEnergyN} calculates it as $1.125$.

\begin{figure*}
\centering
\includegraphics[width=\textwidth]{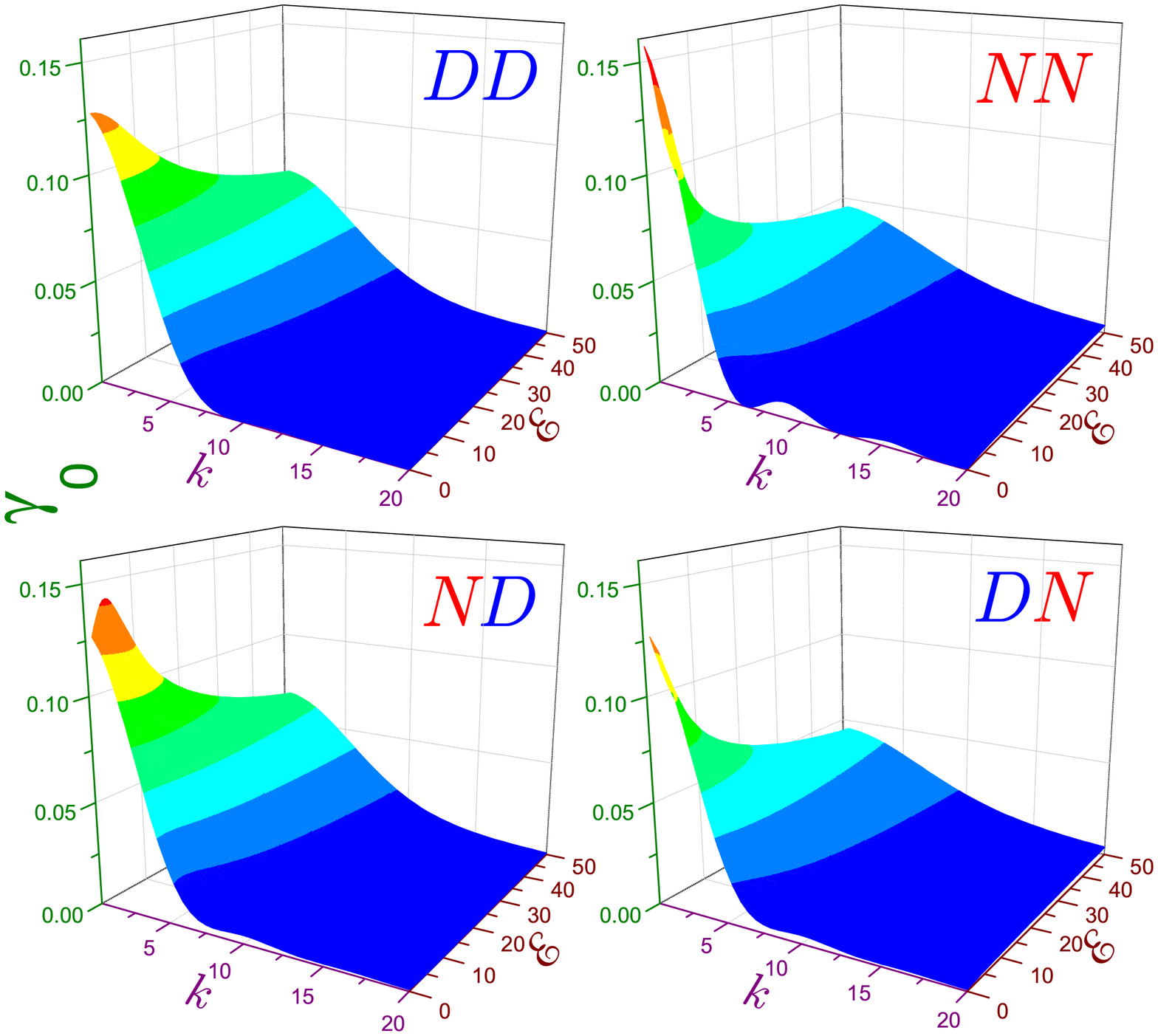}
\caption{\label{MomentumDensities0}
Ground state momentum density $\color{pakistangreen}\gamma_0({\color{sangria}\mathscr{E}};{\color{palatinatepurple}k})$ in terms of the momentum $\color{palatinatepurple}k$ and electric field $\color{sangria}\mathscr{E}$.}
\end{figure*}

\subsection{Entropies}
Ground state momentum density is shown in Fig.~\ref{MomentumDensities0} as a function of the momentum $k$ and voltage $\mathscr{E}$. The general tendency of the electric field influence is the decrease of the maximal values of $\gamma_n$ and smoothing out of the oscillations along the $k$ axis. Note that for the Neumann-Dirichlet well the largest value $\gamma_0(0)$ as  a function of the applied voltage has a local maximum of $~0.1448$ at $\mathscr{E}\sim3$ while for all other geometries it is a monotonically decreasing function of the field. Similar dependencies are characteristic for the higher lying states too.

\begin{figure*}
\centering
\includegraphics[width=\textwidth]{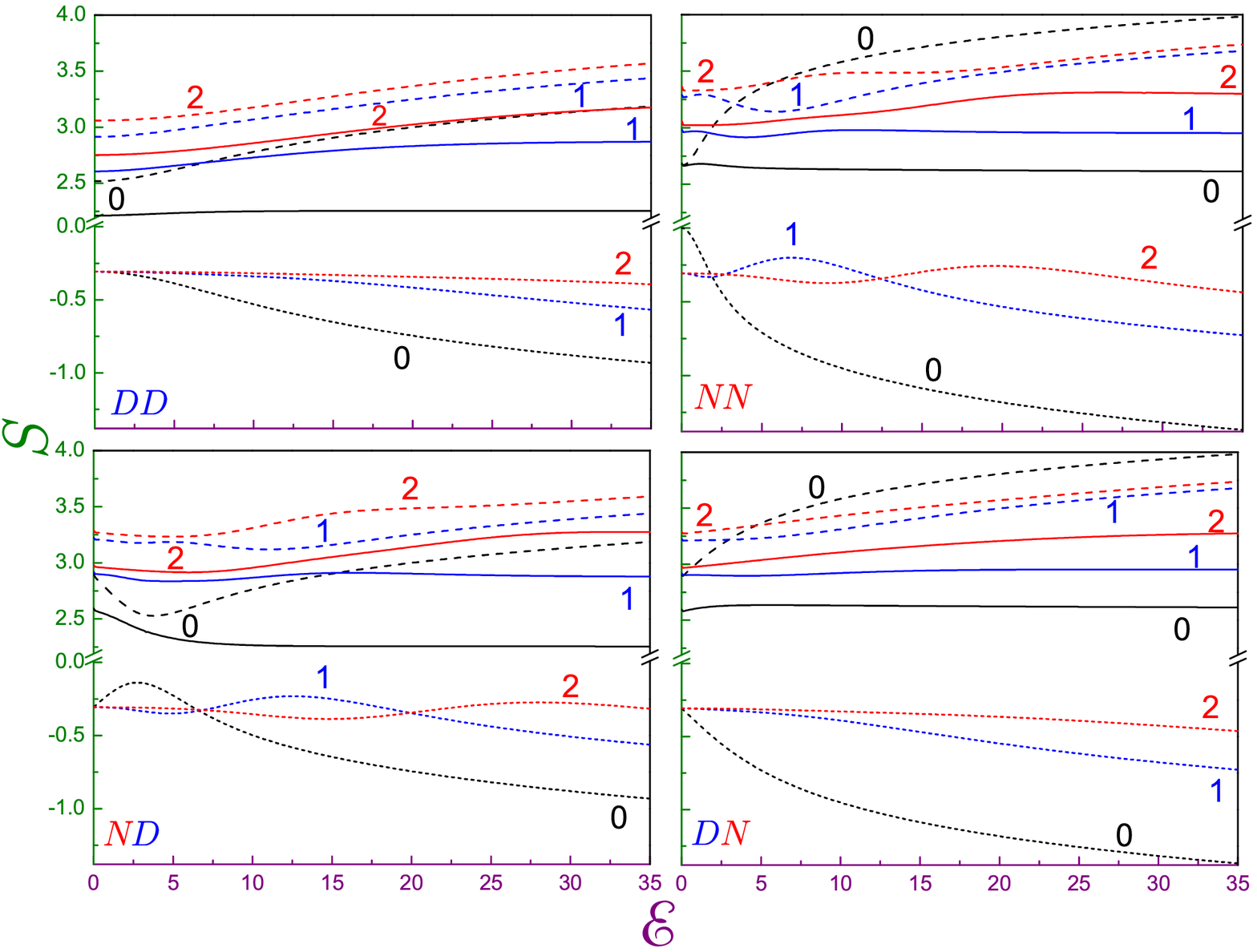}
\caption{\label{Entropies}
Position $\color{pakistangreen}S_x$ (dotted lines), momentum $\color{pakistangreen}S_k$ (dashed lines) and total $\color{pakistangreen}S_t$ (solid curves) entropies as a function of the electric field $\color{palatinatepurple}\mathscr{E}$ for all permutations of the Dirichlet and Neumann BCs. Two characters in the left lower corner of each panel denote a type of the edge requirements while the digits near the curves depict the corresponding quantum number $n$. Note vertical line breaks from $\color{pakistangreen}S=0.01$ to $\color{pakistangreen}S=2.19$.}
\end{figure*}

Peculiar features of the position and momentum densities exemplified by Figs.~\ref{Function1}-\ref{MomentumDensities0} determine the properties of the corresponding entropies $S_x$ and $S_k$.  Fig.~\ref{Entropies} shows them together with the sum $S_t$ as functions of the electric field for all four possible BCs and several low-lying levels. Characteristic feature is a nonmonotonic dependence of the position $S_x$ and momentum $S_k$ entropies on the small and moderate voltages for the structures with the left Neumann wall (except the ground NN level). For example, the maximum of the lowest ND state position entropy is explained by the sharp decrease of the corresponding function $\Psi_0(x)$ at the left surface for $\mathscr{E}\lesssim5$ (see the left lower panel of Fig.~\ref{Function0}) and accompanying decline of its negative contribution to $S_x$ while the corresponding density at the right edge stays almost unchanged. Simultaneously, mentioned above increase of the momentum density results in the contraction of $S_k$ and, as a consequence of these two facts, the lowering of the total entropy that, however, stays above the fundamental limit from the right-hand side of Eq.~\eqref{EntropicInequality1}, as expected. For the higher voltages, the accumulating near the right interface function $\Psi_0$ pushes the position space entropy $S_x$ downward forming the extremum seen in the figure. On the contrary, the Dirichlet left interface subdues the entropy swaying, as the corresponding panels of Fig.~\ref{Entropies} demonstrate. At the large $\mathscr{E}$, the entropies are determined by the right-wall BC, as discussed above; accordingly, they are almost equal for the DN and pure Neumann QWs and their growing with the field magnitudes are larger than their (almost identical) counterparts for the ND and pure Dirichlet structures. Similar to the field-free case, the Dirichlet ground state has the lowest total entropy among all other levels and all possible BC combinations. Note its quite small variation with the voltage: from its zero-field value of $2.2120$ it saturates to approximately $2.2552$ at the large $\mathscr{E}$. Growing with the field magnitude of the negative position entropy $S_x$ means smaller uncertainty in determining particle location while the larger values of the positive momentum entropy $S_k$ indicate that the uncertainty in calculating momentum $k$ increases.

\section{Concluding remarks}\label{sec_Conclusions}
Exact solutions of the Helmholtz equation inside the 1D interval with different BCs at its ends  and subjected to the linear potential inside revealed similarities as well as differences of the miscellaneous permutations of the edge requirements. Similarities include: a continuous growth with the electric field $\mathscr{E}$ of the ground-state polarization for any BC geometry and its slow approach to the saturation value that, however, is different for  the uniform and Zaremba distributions of the surface requirements; a decrease of the excited-level polarization for the small voltages and reverse of its change for the stronger electric forces. The negative polarization at the small $\mathscr{E}$ is conveniently described by the perturbation theory; in particular, it is shown that its reason lies in the field-induced  mixing of the unperturbed levels. Mathematical analysis of the simple two-level approximation proved that only an admixture of the neighbouring states plays a dominant role in the change of the polarization. In addition, the perturbation theory confirms mathematically a physical conclusion about different magnitudes of the polarization for the different BCs. In the opposite limit of the high voltages, physically the quantum particle inside the box does not 'see' one of the surfaces what mathematically means that solutions of the Schr\"{o}dinger equation in this case depend only on the BC at the opposite wall to which it is pushed by the electric field. Application of the Hellmann-Feynman theorem, which long ago proved to be a powerful tool of studying quantum systems \cite{Feynman1,Lowdin1}, in our case revealed a fundamental relation between the speed of the energy change with the field, on the one hand, and the polarization, on the other. Applied voltage also modifies position and momentum densities and entropies in such a way that the entropic uncertainty relation, Eq. ~\eqref{EntropicInequality1}, is always satisfied.

Dirichlet, Eq.~\eqref{Dirichlet1}, and Neumann, Eq.~\eqref{Neumann1}, conditions are the limiting cases of the so called Robin BC \cite{Gustafson1}
\begin{equation}
\left.{\bf n}{\bm\nabla}\Psi\right|_{\cal S}=\left.\frac{1}{\Lambda}\Psi\right|_{\cal S},
\end{equation}
where the variation of the extrapolation length $\Lambda$ continuously changes the edge requirements from the Dirichlet ($\Lambda=0$) to the Neumann ($\Lambda=\infty$) configuration. At the zero voltage, $\mathscr{E}=0$, the vanishingly small negative   Robin length, $\Lambda\rightarrow-0$, heavily pushes two levels to the QW interfaces with their energies unrestrictedly decreasing: $E_{1,2}\xrightarrow[\Lambda\rightarrow-0]{}-1/(\pi\Lambda)^2$ \cite{AlHashimi1,Olendski3}. Their almost full degeneracy in this regime might lead to the nontrivial dependence of the spectrum on the field while  the strong localization at the QW walls might produce unusual polarization and entropy behavior. As a simplifying example, for the single wall with the negative Robin length $\Lambda$ \cite{Seba1,Fulop1,Belchev1,Georgiou1} it is elementary to prove that at $\mathscr{E}=0$ it possesses one bound state with the energy (in regular units)
\begin{equation}\label{SingleRobinWallEnergy}
E=-\frac{\hbar^2}{2m\Lambda^2}
\end{equation}
and the entropies 
\begin{subequations}\label{SingleRobinWallEntropy}
\begin{eqnarray}\label{SingleRobinWallEntropyX}
S_x&=&1-\ln2+\ln|\Lambda|\\
\label{SingleRobinWallEntropyK}
S_k&=&2\ln2+\ln\pi-\ln|\Lambda|,
\end{eqnarray}
\end{subequations}
where it is assumed that $\hbar^2/(2m)\equiv1$. Eqs.~\eqref{SingleRobinWallEnergy} and \eqref{SingleRobinWallEntropy} are exact for any negative $\Lambda$. It is seen from them that even though the position and momentum entropies diverge at the limiting values of $\Lambda\rightarrow-\infty$ and $\Lambda\rightarrow-0$, the total entropy
\begin{equation}
S_t=1+\ln\pi+\ln2
\end{equation}
stays finite, does not depend on the extrapolation length (due to the exact cancelling of the divergent terms in $S_x$ and $S_k$) and does satisfy inequality \eqref{EntropicInequality1}.

At the weak fields the polarization for any type of the BCs has opposite signs for the ground and any excited states. The natural question arises: Is there ``warm" enough temperature or/and sufficient number of the charged fermions that are able to push the total statistically averaged polarization to the negative values? The following paper \cite{Olendski2}, among other issues, answers this question.

\section{Acknowledgement}\label{sec_6}
This project was supported by Deanship of Scientific Research, College of Science Research Center, King Saud University.


\begin{thebibliography}{100}
\bibitem{Landau1}L. D. Landau and E. M. Lifshitz, Quantum Mechanics (Non-Relativistic Theory) (Pergamon, New York, 1977).
\bibitem{Rabinovitch1}A. Rabinovitch and J. Zak, \PRB {\bf 4}, 2358 (1971).
\bibitem{Lukes1}T. Lukes, G. A. Ringwood, and B. Suprapto, \PA {\bf 84}, 421 (1976).
\bibitem{Fernandez1}F. M. Fern\'{a}ndez and E. A. Castro, \PA {\bf 111}, 334 (1982).
\bibitem{Bastard1}G. Bastard, E. E. Mendez, L. L. Chang, and L. Esaki, \PRB {\bf 28}, 3241 (1983).
\bibitem{Ahn1}D. Ahn and S. L. Chuang, \APL {\bf 49}, 1450 (1986).
\bibitem{Nguyen1}D. Nguyen and T. Odagaki, \AJP {\bf 55}, 466 (1987).
\bibitem{Miller1}D. A. B. Miller, D. S. Chemla, T. C. Damen, A. C. Gossard, W. Wiegmann, T. H. Wood, and C. A. Burrus, \PRL {\bf 53}, 2173 (1984).
\bibitem{Miller2}D. A. B. Miller, D. S. Chemla, T. C. Damen, A. C. Gossard, W. Wiegmann, T. H. Wood, and C. A. Burrus, \PRB {\bf 32}, 1043 (1985).
\bibitem{Weiner1}J. S. Weiner, D. A. B. Miller, D. S. Chemla, T. C. Damen, C. A. Burrus, T. H. Wood, A. C. Gossard, and W. Wiegmann, \APL {\bf 47}, 1148 (1985).
\bibitem{Miller3}D. A. B. Miller, D. S. Chemla, and S. Schmitt-Rink, \PRB {\bf 33}, 6976 (1986).
\bibitem{Achtstein1}A W. Achtstein, A. V. Prudnikau, M. V. Ermolenko, L. I. Gurinovich, S. V. Gaponenko, U. Woggon, A. V. Baranov, M. Y. Leonov, I. D. Rukhlenko, A. V. Fedorov, and M. V. Artemyev, ACS Nano {\bf 8}, 7678 (2014).
\bibitem{Wood1}T. H. Wood, C. A. Burrus, D. A. B. Miller, D. S. Chemla, T. C. Damen, A. C. Gossard, and W. Wiegmann, IEEE J. Quantum Electron. {\bf 21}, 117 (1985).
\bibitem{Miller4}D. A. B. Miller, D. S. Chemla, T. C. Damen, T. H. Wood, C. A. Burrus, A. C. Gossard, and W. Wiegmann, IEEE J. Quantum Electron. {\bf 21}, 1462 (1985).
\bibitem{Wood2}T. H. Wood, C. A. Burrus, A. H. Gnauck, J. M. Wiesenfeld, D. A. B. Miller, D. S. Chemla, and T. C. Damen, \APL {\bf 47}, 190 (1985).
\bibitem{Jackson1}J. D. Jackson, Classical Electrodynamics, 3rd edn. (Wiley, New York, 1999).
\bibitem{DeGennes1}P. G. de Gennes, Superconductivity of Metals and Alloys (Benjamin, New York, 1966).
\bibitem{Olendski1}O. Olendski and L. Mikhailovska, \PRE {\bf 67}, 056625 (2003).
\bibitem{Davies1}K. Davies, Ionospheric Radio (Peter Peregrinus, London, 1990).
\bibitem{Budden1}K. G. Budden, The Wave-Guide Mode Theory of Wave Propagation (Prentice-Hall, Englewood Cliffs, N. J., 1961).
\bibitem{Johnson1}E. R. Johnson, M. Levitin, and L. Parnovski, SIAM J. Math. Anal. {\bf 37}, 1465 (2006).
\bibitem{Zaremba1}S. Zaremba, J. Math. Pure Appl. 9 S\`{e}r. {\bf 6}, 127 (1927).
\bibitem{Najar1}H. Najar and O. Olendski, \jpa {\bf 44}, 305304 (2011).
\bibitem{Exner1}P. Exner, \JPA {\bf 28}, 5323 (1995).
\bibitem{Budden2}K. G. Budden, The Propagation of Radio Waves (Cambridge, Cambridge, 1985).
\bibitem{Gaponenko1}S. V. Gaponenko, Introduction to Nanophotonics  (Cambridge, Cambridge, 2010).
\bibitem{Grebenkov1}D. S. Grebenkov and B.-T. Nguyen, SIAM Rev. {\bf 55}, 601 (2013).
\bibitem{Olendski2}O. Olendski, Ann. Phys. (Berlin) {\bf 527}, 296 (2015).
\bibitem{Abramowitz1}M. Abramowitz and I. A. Stegun, Handbook of Mathematical Functions (Dover, New York, 1964).
\bibitem{Vallee1}O. Vall\'{e}e and M. Soares, Airy Functions and Applications to Physics, 2nd edn (World Scientific, Hackensack, NJ, 2010).
\bibitem{Bialynicki1}I. Bia{\l}ynicki-Birula, \PLA {\bf 103}, 253 (1984).
\bibitem{Everett1}H. Everett, III, in: The Many-Worlds Interpretation of Quantum Mechanics, edited by B. S. DeWitt and N. Graham, Princeton Series in Physics (Princeton University Press, Princeton, 1973), chap. 1.
\bibitem{Hirschman1}I.I. Hirschman, \AJM {\bf 79}, 152 (1957).
\bibitem{Bialynicki2}I. Bia{\l}ynicki-Birula and J. Mycielski, \CMP {\bf 44}, 129 (1975).
\bibitem{Beckner1}W. Beckner, \AM {\bf 102}, 159 (1975).
\bibitem{Bialynicki3}I. Bia{\l}ynicki-Birula and {\L}. Rudnicki, in: Statistical Complexity: Applications in Electronic Structure, edited by K. D. Sen (Springer, Dordrecht, 2011), chap. 1.
\bibitem{Wehner1}S. Wehner and A. Winter, New J. Phys. {\bf 12}, 025009 (2010).
\bibitem{Bialynicki4}I. Bia{\l}ynicki-Birula, AIP Conf. Proc. {\bf 889}, 52 (2007).
\bibitem{AlHashimi1} M. H. Al-Hashimi and U.-J. Wiese, \APNY {\bf 327}, 1 (2012).
\bibitem{Gadre1}S. R. Gadre, S. B. Sears, S. J. Chakravorty, and R. D. Bendale, \PRA {\bf 32}, 2602 (1985).
\bibitem{Robinett1}R. W. Robinett, \AJP {\bf 63}, 823 (1995).
\bibitem{Tao1}J. Tao, G. Li, and J. Li, and J. Li, J. Chem. Phys. {\bf 107}, 1227 (1997).
\bibitem{SanchezRuiz1}J. S\'{a}nchez-Ruiz, \PLA {\bf 226}, 7 (1997).
\bibitem{Majernik1}V. Majern\'{i}k and L. Richterek, \JPA {\bf 30}, L49 (1997).
\bibitem{Majernik2}V. Majern\'{i}k and L. Richterek, Eur. J. Phys. {\bf 18}, 79 (1997).
\bibitem{Majernik3}V. Majern\'{i}k, R. Charvot, and E. Majern\'{i}kov\'{a}, \JPA {\bf 32}, 2207 (1999).
\bibitem{Dehesa1}J. S. Dehesa, A. Mart\'{i}nez-Finkelshtein, and J. S\'{a}nchez-Ruiz, J. Comp. Appl. Math. {\bf 133}, 23 (2001).
\bibitem{LopezRosa1}S. L\'{o}pez-Rosa, J. Montero, P. S\'{a}nchez-Moreno, J. Venegas, and J. S. Dehesa, J. Math. Chem. {\bf 49}, 971 (2011).
\bibitem{Aptekarev1}A. I. Aptekarev, J. S. Dehesa, P. S\'{a}nchez-Moreno, and D. N. Tulyakov, J. Math. Chem. {\bf 50}, 1079 (2012).
\bibitem{Laguna1}H. G. Laguna and R. P. Sagar, \APB {\bf 526}, 555 (2014).
\bibitem{Prudnikov1}A. P. Prudnikov, Y. A. Brychkov, and O. I. Marichev, Integrals and Series  Vol. 1 (Gordon and Breach, New York, 1986), p. 540, Eq.~2.6.34.12.
\bibitem{Prudnikov2_1}A. P. Prudnikov, Y. A. Brychkov, and O. I. Marichev, Integrals and Series  Vol. 2 (Gordon and Breach, New York, 1990).
\bibitem{Prudnikov3_1}A. P. Prudnikov, Y. A. Brychkov, and O. I. Marichev, Integrals and Series  Vol. 3 (Gordon and Breach, New York, 1990).
\bibitem{Brychkov1}Y. A. Brychkov, Handbook of Special Functions: Derivatives, Integrals, Series and Other Formulas (Chapman \& Hall, Boca Raton, 2008).
\bibitem{Gradshteyn1}I. S. Gradshteyn and I. M. Ryzhik, Table of Integrals, Series, and Products (Academic, New York, 2014).
\bibitem{Prudnikov1_1}Ref.~\cite{Prudnikov1}, p. 687.
\bibitem{Prudnikov2}Ref.~\cite{Prudnikov1}, p. 685, Eq. 5.1.25.4.
\bibitem{Prudnikov3}Ref.~\cite{Prudnikov1}, p. 688, Eq. 5.1.26.1.
\bibitem{Prudnikov4}Ref.~\cite{Prudnikov1}, p. 653, Eq. 5.1.4.1.
\bibitem{Feynman1}R. P. Feynman, \PR {\bf 56}, 340 (1939).
\bibitem{Lowdin1}P.-O. L\"{o}wdin, J. Mol. Spectrosc. {\bf 3}, 46 (1959).
\bibitem{Montgomery1}H. E. Montgomery, Jr. and V. I. Pupyshev, Eur. Phys. J. H {\bf 38}, 519 (2013).
\bibitem{Lipson1}A. Lipson, S. G. Lipson and H. Lipson, Optical Physics, 4th edn. (Cambridge, Cambridge, 2011).
\bibitem{Rudin1}W. Rudin, Principles of Mathematical Analysis, 3rd edn. (McGraw-Hill, New-York , 1976).
\bibitem{Katriel1}J. Katriel and G. Adam, Physica {\bf 43}, 546 (1969).
\bibitem{Maple1}http://www.maplesoft.com
\bibitem{Gustafson1}K. Gustafson and T. Abe, Math. Intell. {\bf 20}({\em 1}), 63 (1998).
\bibitem{Olendski3}O. Olendski and L. Mikhailovska, \PRE {\bf 81}, 036606 (2010).
\bibitem{Seba1}P. \v{S}eba, Lett. Math. Phys. {\bf 10}, 21 (1985).
\bibitem{Fulop1}T. F\"{u}l\"{o}p, T. Cheon, and I. Tsutsui, \PRA {\bf 66}, 052102 (2002).
\bibitem{Belchev1}B. Belchev and M. A. Walton, \jpa {\bf 43}, 085301 (2010).
\bibitem{Georgiou1}O. Georgiou, G. Gligori\'{c}, A. Lazarides, D. F. M. Oliveira, J. D. Bodyfelt, and A. Goussev, Europhys. Lett. {\bf 100}, 20005 (2012).
\end{thebibliography}
\end{document}